\documentclass[aps,twocolumn,pre, superscriptaddress]{revtex4-2}

\usepackage[english]{babel}
\usepackage{amsmath, amssymb, mathtools, physics}
\usepackage{braket}
\usepackage{ifthen}
\usepackage{graphicx,float}
\usepackage{multirow}
\usepackage[RGB,dvipsnames]{xcolor}
\definecolor{dodgerblue}{rgb}{0.12, 0.56, 1.0}
\definecolor{darkcyan32144140}{RGB}{32,144,140}
\definecolor{darkslateblue5294141}{RGB}{52,94,141}      
\definecolor{darkslateblue6467135}{RGB}{64,67,135}
\definecolor{darkslateblue7235116}{RGB}{72,35,116}      
\definecolor{darkslategray38}{RGB}{38,38,38}
\definecolor{greenyellow18922238}{RGB}{189,222,38}      
\definecolor{indigo68184}{RGB}{68,1,84}                 
\definecolor{lavender234234242}{RGB}{234,234,242}
\definecolor{mediumseagreen34167132}{RGB}{34,167,132}   
\definecolor{mediumseagreen68190112}{RGB}{68,190,112}
\definecolor{teal41120142}{RGB}{41,120,142}
\definecolor{yellowgreen12120981}{RGB}{121,209,81}
\colorlet{bluedarkslateblue5294141}{blue!50!darkslateblue5294141}

\newcommand{\Path}[1]{\mathcal{P} \ifthenelse{\equal{#1}{}}{}{[#1]}}                    
\newcommand{\Pathr}[1]{\widetilde{\mathcal{P}}\ifthenelse{\equal{#1}{}}{}{[#1]}}        
\newcommand{\rev}[1]{\widetilde{#1}}                                                    
\newcommand{\R}{\mathcal{R}} 

\newcommand{\mean}[1]{\left\langle #1\right\rangle}

\newcommand{\wtdPsi}[3]{\Psi_{#1\to #2}(#3)}
\newcommand{\psteady}[1]{p_{#1}^\text{s}}
\newcommand{\jsteady}[1]{j_{#1}^\text{s}}
\newcommand{\meanNness}[1]{\left\langle\dot{n}_{#1}\right\rangle_\text{s}}
\newcommand{\NiIJ}[2]{N_{#1,#2}}
\newcommand{\MiIJK}[3]{N_{#1,#2\setminus #3}}
\newcommand{\MiIJKedges}[2]{#1\setminus #2}
\newcommand{\Psiaff}[2]{a_{#1}(#2)}
\newcommand{\PsiaffLn}[4]{\ln\frac{\wtdPsi{#1}{#2}{t}}{\wtdPsi{#3}{#4}{t}}}
\newcommand{\spath}[2]{\hat{a}_{#1}(#2)}
\newcommand{\obsNtwo}[1]{u_{#1}}
\newcommand{\obsMtwoK}[2]{u_{#1\setminus #2}}
\usepackage{hyperref}
\usepackage[nameinlink]{cleveref}
\usepackage{enumitem}
\usepackage{tikz}
\usetikzlibrary{external,cd}
\usepackage{pgfplots}
\usetikzlibrary{automata, arrows.meta, positioning, calc,shapes}
\usetikzlibrary{nfold}
\makeatletter
\tikzset{
  side by side/.style args={#1:#2}{
    postaction={path only,draw=#1,offset=+.5\pgflinewidth},
    postaction={path only,draw=#2,offset=+-.5\pgflinewidth}},
  side by side3/.style args={#1:#2:#3}{
    postaction={path only,draw=#1,offset=+1.0\pgflinewidth},
    postaction={path only,draw=#2,offset=+.0\pgflinewidth},
    postaction={path only,draw=#3,offset=+-1.0\pgflinewidth}},
  side by side'/.style={path only,side by side={#1}},
  offset/.code=
    \tikz@addoption{
      \pgfgetpath\tikz@temp
      \pgfsetpath\pgfutil@empty
      \pgfoffsetpath\tikz@temp{#1}}}
\makeatother
\tikzset{
  subroutine/.style={
    state,
    rectangle,
    draw,
    inner sep=6pt,
    append after command={
    \pgfextra{
      \draw ([xshift=0.3em]\tikzlastnode.north west) to ([xshift=0.3em]\tikzlastnode.south west);
      \draw ([xshift=-0.3em]\tikzlastnode.north east) to ([xshift=-0.3em]\tikzlastnode.south east);
      }
    }
  },
  inputs/.style={
    state,
    trapezium,
    trapezium left angle=75,
    trapezium right angle=105,
    draw,
    inner sep=6pt
  },
  process/.style={
    state,
    rectangle,
    draw,
    inner sep=6pt
  },
  result/.style={
    state,
    trapezium,
    trapezium left angle=105,
    trapezium right angle=105,
    draw,
    inner sep=6pt
  },
  functions/.style={
    state,
    rectangle,
    rounded corners=3pt,
    draw,
    inner sep=6pt
  },
  decision/.style={
    state,
    diamond,
    aspect=2,
    draw,
    inner sep=2pt
  },
  startEnd/.style={
    state,
    rectangle,
    draw=none,
    inner sep=6pt,
    overlay,
    append after command={
      \pgfextra{
        \draw ([xshift=0.2em]\tikzlastnode.south west) to ([xshift=-0.2em]\tikzlastnode.south east)
        to[out=0, in=0] ([xshift=-0.2em]\tikzlastnode.north east) to ([xshift=0.2em]\tikzlastnode.north west)
        to[out=180, in=180] ([xshift=0.2em]\tikzlastnode.south west);
      }
    }
  }
}
\usepackage{scalerel,stackengine}
\newcommand\correspondshat{\mathrel{\stackon[1.5pt]{=}{\stretchto{%
  \scalerel*[\widthof{=}]{\wedge}{\rule{1ex}{3ex}}%
  }{0.5ex}}}}
\begin{document}
%
\title{From observed transitions to hidden paths in Markov networks}

\author{Alexander M. Maier}
\author{Udo Seifert}%
\affiliation{%
 II. Institut für Theoretische Physik, Universität Stuttgart, 70550 Stuttgart, Germany
}%
\author{Jann van der Meer}%
\affiliation{%
 II. Institut für Theoretische Physik, Universität Stuttgart, 70550 Stuttgart, Germany
}%
\affiliation{%
Department of Physics No. 1, Graduate School of Science, Kyoto University, Kyoto 606-8502, Japan
}%
\date{\today}

\begin{abstract}
The number of observable degrees of freedom is typically limited in experiments. Here, we consider discrete
Markov networks in which an observer has access to a few visible transitions and the waiting times between 
these transitions. Focusing on the underlying structure of a discrete network, we present methods to infer 
local and global properties of the network from observed data. First, we derive bounds on the microscopic 
entropy production along the hidden paths between two visible transitions, which complement extant bounds
on mean entropy production and affinities of hidden cycles. Second, we demonstrate how the operationally 
accessible data encodes information about the topology of shortest hidden paths, which can be used to 
identify potential clusters of states or exclude their existence. Finally, we outline a systematic way to 
combine the inferred data, resulting in an algorithm that finds the candidates for a minimal graph of the 
underlying network, i.e., a graph that is part of the original one and compatible with the observations. 
Our results highlight the interplay between thermodynamic methods, waiting-time distributions and 
topological aspects like network structure, which can be expected to provide novel insights in 
other set-ups of coarse graining as well.
\end{abstract}
\maketitle

\section{Introduction}\label{sec:intro}

On the microscopic scale, the laws of thermodynamics acquire a form that is different to the expressions commonly used for macroscopic descriptions. As thermal noise significantly affects the trajectory of a given physical system embedded in a thermal environment, we have to resort to a stochastic rather than deterministic description of the dynamics. The framework of stochastic thermodynamics adopts this conceptual shift for thermodynamic quantities such as entropy production, work, and heat, which are now understood as trajectory-dependent random variables \cite{seki10, jarz11, seif12, peli21, shir23}. As a result, the thermodynamic properties of a physical system are now closely related to its stochastic dynamics, so that -- given the right tools -- we are not only able to incorporate the thermodynamic laws into a physical model but also to investigate the reverse problem, namely the inference of dynamical aspects of a given physical system from its thermodynamic characteristics. 

Such questions are studied in the emerging field of thermodynamic inference \cite{seif19}, which focuses on physical systems that can be observed only partially. In such a situation, the principles of stochastic thermodynamics cannot be applied directly, and the development of appropriate tools for inference constitutes one of the main tasks of this applied branch of stochastic thermodynamics. Much of the recent research in this field can be roughly sorted into one of two categories. A substantial number of studies focus on a specific nonequilibrium phenomenon and quantify the amount of dissipation implied by such an observation. Typically formulated as a lower bound on a quantifier for nonequilibrium like entropy production, such results include, e.g., the preeminent thermodynamic uncertainty relation \cite{seif15, ging16, horo20}, speed limits and transport results \cite{shir18, ito20, dech22a, vu23a, naga25}, response inequalities \cite{dech20, owen20, asly24}, bounds on the coherence of oscillations \cite{ohga23, shir23b} and bounds in terms of correlation or waiting times \cite{skin21a, dech23, piet24}. 

This first category of results is complemented by a second one that places more emphasis on how the coarse-grained dynamics, i.e., the dynamics at the observable level, relates to the thermodynamics on the microscopic level. A rough classification of studies in this direction can be made by distinguishing the assumed observables and the mathematical structure of the coarse-grained process. Older works mainly focus on state lumping or decimation techniques, which result in phenomenological equations for the dynamics \cite{espo12, bo17, seif19}. Over time, attention gradually shifted from the states to the currents along observed edges connecting them \cite{shir15, pole19, mart19}, a development that culminated in a thermodynamically consistent framework for coarse graining based on the transitions themselves \cite{vdm22, haru22}. One reason for this shift is that more powerful technical tools can be employed to connect dynamics and thermodynamics. Whereas in the case of state lumping typically only lower bounds on the mean entropy production rate expressed as a Kullback-Leibler divergence are possible \cite{gome08a, rold10, muy13}, the more recent transition-based description allows for stronger results on the trajectory level, which include fluctuation relations \cite{vdm22, haru23a} and a meaningful identification of a coarse-grained entropy production \cite{degu24, degu24b} even in the presence of time-dependent driving. Conceptually, these methods are able to include waiting times in their framework, a trait that is shared by related coarse-graining schemes that rely on milestoning \cite{hart21, blom24a}, Markovian events \cite{vdm23, degu24, degu24b} and trimming \cite{igos25a}. 

\begin{figure*}
\centering
\includegraphics[scale=1]{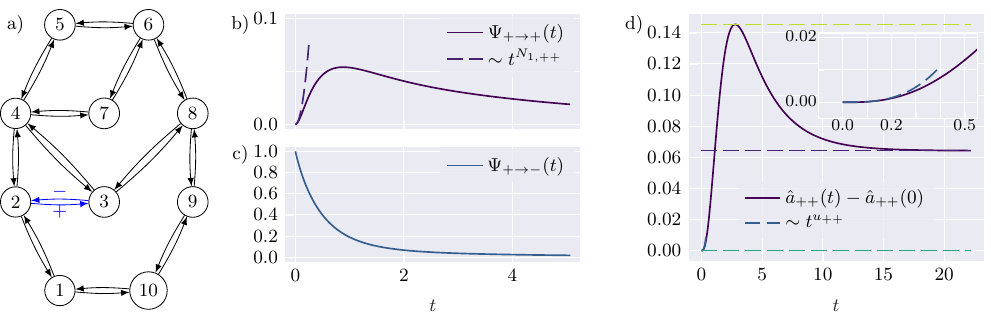}
\caption{Waiting-time distributions in a multicyclic Markov network. a) Graph of a network, 
with transition rates given in \Cref{app:MPSeggtopo}. Blue edges $\pm$ are visible while other edges and states are hidden.
b) Waiting-time distribution $\wtdPsi{+}{+}{t}$ for two consecutive $+$ transitions. The distribution is proportional to
$t^{\NiIJ{1}{++}}$ in the short-time limit, which yields the number $\NiIJ{1}{++} =2$ of hidden transitions along the shortest cycle containing
$\pm$. c) Waiting-time distribution $\wtdPsi{+}{-}{t}$ of the observed sequence $+\to -$. The value
$\wtdPsi{+}{-}{0}$ yields the rate of the transition $-$.
d) The difference $\spath{++}{t} - \spath{++}{0}$, which enters $\obsNtwo{++}$ in \Cref{eq:NtwoIJ}. Its
short-time limit is proportional to $t^{\obsNtwo{++}}$ with $\obsNtwo{++} = \NiIJ{2}{++} - \NiIJ{1}{++} = 3$. The horizontal dashed lines mark local extrema of
$\spath{++}{t} - \spath{++}{0}$. The global extrema yield bounds on cycle affinities as we will discuss in \Cref{sec:INPpbeEstimation}.
} \label{fig:setupExamples}
\end{figure*}
Apart from studying specific nonequilibrium phenomena or coarse-grained models, we might speculate about a third direction of inference that attempts to uncover not only thermodynamic but also dynamical or topological features of the underlying microscopic model. In this work, we assume that the underlying dynamics takes place in a discrete, finite configuration space and is described by a master equation, which is a common assumption within stochastic thermodynamics. We study the scenario that an external observer can register transitions between some pairs of states, whereas the states themselves and the remaining transitions remain inaccessible. In this setting, which is further detailed in \Cref{sec:setup}, we propose a number of novel approaches that reveal local and global properties of the underlying microscopic model, thereby extending the transition-based methods introduced in Refs. \cite{vdm22, haru22}.

A common theme in the methods introduced in this work is the focus on microscopic paths, i.e., the possible trajectories in the hidden part of the system that take place between two observed transitions. One of our main results is a constructive statement regarding the existence of microscopic paths between particular transitions whose entropy production is larger or smaller than a particular, operationally determinable threshold. We provide the precise formulation of this statement and a numerical demonstration in \Cref{sec:INPpbeEstimation}. The remainder of our results is built around a systematic connection between distributions of the waiting time between successive transitions and the minimal number of hidden transitions that take place in between, which is detailed in \Cref{sec:shortestpathsL}. Although this technique itself is not new (see, e.g., Refs. \cite{li13, vdm22} for states and transitions, respectively), we present two novel refinements that focus on structural properties of the network on which the dynamics takes place. First, such structural information may provide insights into whether a network contains distinct clusters, a hypothesis that will be formulated more rigorously in \Cref{sec:INPwsfclusters}. Second, we may speculate about whether a systematic way of combining the inferable topological information is sufficient to reconstruct the entire microscopic model or a minimal variant of it. Our proposal towards this challenging goal is presented as an algorithm in \Cref{sec:constrMinGraph}. We discuss our methods and their limitations in \Cref{sec:discussion} before concluding with an outlook on potential future work in \Cref{sec:outlook}.

\section{Setup}\label{sec:setup}
We consider a connected network of $N$ states $i\in\lbrace 1,2,\dots N\rbrace$. The state $i(t)$ of the
system at time $t$ follows a stochastic dynamics with instantaneous transitions between states that
share an edge in the underlying graph. Transitions from state $i$ to $j$ occur according to a time-independent rate $k_{ij}$,
which fulfills the local detailed balance condition $k_{ij}/k_{ji}=\exp(F_i - F_j + f_{ij})$ with free
energy $F_i$ of state $i$ and non-conservative contribution $f_{ij}=-f_{ji}$ in units of thermal energy.
Hence, in the graph edges come in pairs, which we will call links. 
Moreover, the rates $k_{ij}$ generate the dynamics of the occupation probability $p_i(t)$ of state $i$ at
time $t$, which obeys the master equation
\begin{align}
\partial_t p_i(t) = \sum_{j} \left[p_j(t)k_{ji} - p_i(t)k_{ij}\right].
\end{align}
In the long-time limit $t\to\infty$, $p_i(t)$ reaches a stationary state $\psteady{i}$ with constant net
current $\jsteady{ij} = \psteady{i}k_{ij} - \psteady{j}k_{ji}$ through the link from $i$ to $j$.
We assume that the network has reached this steady state throughout the paper.

We assume an observer for whom the underlying Markovian dynamics of the network is only partially accessible.
In particular, the observer can register passages along $m$ pairs of edges and can distinguish forward from backward transitions.
In the steady state, the rate of observed transitions $I= (ij)$ is given by $P(I) \equiv \psteady{i}k_{ij}$. 
The remaining transitions and the states of the network are hidden from the observer, as illustrated in
\Cref{fig:setupExamples}\,a) for the case $m = 1$. With access to a few transitions only, the observer
aims at learning about the topology and the kinetics of the partially hidden network from measured data. 

This data characterizes the registered effective dynamics through waiting-time distributions of the form \cite{vdm22, haru22}
\begin{align}
\wtdPsi{I}{J}{t} \equiv P\left(J,t\middle\vert I,t_0=0\right) = k_{kl}p_{k}(t\vert j,0), \label{eq:defwtdPsi}
\end{align}
where $I=(ij)$, $J=(kl)$ and $t$ is the time since registering transition $I$. The function $\wtdPsi{I}{J}{t}$ characterizes the 
probability distribution of registering transition $J$ after waiting time $t$ given that $I$ was observed
at $t_0 = 0$. Its normalization is $\sum_J \int_{0}^{\infty} \wtdPsi{I}{J}{t} \dd{t} = 1$.  Since the underlying state of the network
immediately before a transition is unique, the Markov property of the underlying dynamics ensures that the statistics entering $\wtdPsi{I}{J}{t}$
does not depend on the past prior to $I$. \par

Next, using the reverse transitions $\tilde{I}=(ji),\tilde{J}=(lk)$ of $I$ and $J$, we define
\begin{align}
\spath{IJ}{t} \equiv \ln\frac{P(I)}{P(J)} + \ln\frac{\wtdPsi{I}{J}{t}}{\wtdPsi{\tilde{J}}{\tilde{I}}{t}}. \label{eq:defaIJ}
\end{align}
This quantity is time-independent when there is only one self-avoiding path between the two visible transitions, such as in a unicyclic
network with one visible link. This time-independence of $\spath{IJ}{t}$ is a manifestation of a more general symmetry in waiting-time
distributions between forward and backward paths that holds true as long as there are no corresponding underlying paths that include a
part of a cycle with nonzero affinity \cite{bere06, bere19, vdm22, degu24}. The same quantity also qualifies for an interpretation as
an expression for entropy production that fluctuates on the coarse-grained level \cite{degu24, degu24b}.

\section{Bound on entropy production of microscopic paths}\label{sec:INPpbeEstimation}

\begin{figure*}
\centering
\includegraphics[scale=1]{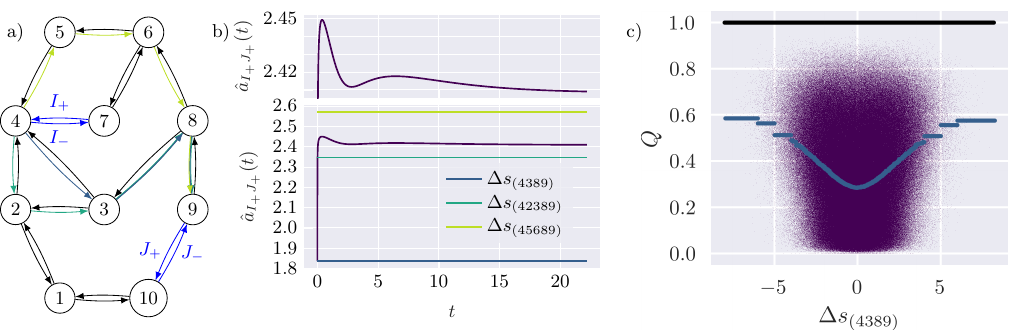}
\caption{Entropy production along paths in a multicyclic 10-state Markov network. 
a) Graph with highlighted paths between visible transitions of the network with transition rates given in \Cref{app:MPSeggpathepr}.
Only transitions $I_\pm,J_\pm$ along the blue edges are visible, whereas states and the remaining transitions are hidden. Three different possible paths between states $4$ and $9$, i.e., between the transitions $I_+$ to $J_+$ are highlighted: The shortest connection along $(4389) = 4 \to 3 \to 8 \to 9$ is marked in blue, whereas the paths $(42389)$ and $(45689)$ are highlighted in teal and olive green, respectively.
b) Time-dependence of $\spath{I_+J_+}{t}$ as the waiting time $t$ between the initial transition $I_+$ and the final transition $J_+$ is varied (purple curve) compared to the entropy production $\Delta s$ of three different possible microscopic paths from $I_+$ to $J_+$ (horizontal lines). Except for paths containing closed loops, these three paths are the only possible ones from $I_+$ to $J_+$. Therefore, the result \eqref{eq:sec3:aff_result} implies that $\spath{I_+J_+}{t}$ stays bounded between the highest and lowest horizontal line for all $t$. In the figure, the value of the lower bound is attained in the limit $t \to 0$, because $\lim_{t \to 0} \spath{I_+J_+}{t}$ precisely recovers the entropy production of the path with the lowest number of intermediate states if this path is unique.
c) Illustration of the quality of the bound \eqref{eq:sec3:aff_result}. We depict the quality factor $Q = \sup_t\vert\spath{I_+J_+}{t} - \Delta s_{(4389)}\vert/\max_{i\in\lbrace 1,2,3
\rbrace}\vert\Delta s_{\gamma_i} - \Delta s_{(4389)}\vert$ for the estimator $\max_t\vert\spath{I_+J_+}{t} - \Delta s_{(4389)}\vert$, which satisfies $0 \leq Q \leq 1$ by virtue of \Cref{eq:sec3:aff_result}, as indicated by the horizontal line. The value $Q$ is calculated for $1 731 699$ randomly selected configurations of transition rates, more details are given in \Cref{app:MPSeggpathepr} (purple dots). The average performance of the quality factor ranges from $Q \simeq 0.3$ to $Q \simeq 0.6$ and is measured as a binned mean grouping similar values of $\Delta s_{(42389)}$  (blue lines).
}
\label{fig:PBEEexample}
\end{figure*}

Our first result combines thermodynamic and topological reasoning to conclude the existence of particular microscopic paths (without self-crossings) whose entropy production exceeds or falls below a certain threshold. It is worth noting that the bound involves the entropy production of an individual microscopic path rather than the average dissipation of a class of paths or trajectories that become indistinguishable after coarse graining. We limit this section to the presentation, discussion and illustration of the result and refer to \Cref{sec:app:affinity_proof} for the proof.

\subsection{Result and discussion}

It is sensible to speculate whether we can attribute thermodynamic meaning to the ratio of waiting-time distributions defined in Eq. \eqref{eq:defaIJ}. Here, we present an interpretation of the quantity $\spath{IJ}{t}$ as a bound on the extremal microscopic entropy production along direct paths $\gamma$ connecting the transition $I$ to $J$. As discussed in more detail below, this perspective complements extant relations establishing $\spath{IJ}{t}$ as a lower bound on the average dissipation of all paths from $I$ to $J$ \cite{degu24} and as an estimator for cycle affinities in the case $I = J$ \cite{vdm22}. 

The novel bound introduced in this work and proved in \Cref{sec:app:affinity_proof} links the extremal values of $\spath{IJ}{t}$ as a function of $t$ to the microscopic entropy production $\Delta s[\hat{\gamma}]$ of particular paths $\hat{\gamma}$ in the form
\begin{equation}
    \min_{\gamma|I \to J} \Delta s[\hat{\gamma}] \leq \inf_t \spath{IJ}{t} \leq \sup_t \spath{IJ}{t} \leq \max_{\gamma|I \to J} \Delta s[\hat{\gamma}] 
    \label{eq:sec3:aff_result}
.\end{equation}
The maximum and minimum are taken over all possible microscopic paths $\gamma$ that start with the transition $I$, end with the transition $J$ and do not contain any other observed transitions. The entropy production $\Delta s$ is, however, evaluated for the trimmed path $\hat{\gamma}$, which is obtained from $\gamma$ by removing all closed loops from the path. The term $\Delta s[\hat{\gamma}]$ includes the contribution to entropy production due to the initial jump $I$ but excludes the last jump $J$. As an example, for the network depicted in Fig. \ref{fig:PBEEexample} and the transitions $I = 7 \to 4$ and $J = 9 \to 10$, a possible microscopic path $\gamma$ and its corresponding trimmed counterpart $\hat{\gamma}$ read
\begin{alignat}{10}
\gamma = \overset{I}{\to} 4 \to 3 & \to 2 \to 4 \to 3 && \to 8 \to 6 \to 8 && \to 9 \overset{J}{\to} \text{ and } \nonumber \\
\hat{\gamma} = \overset{I}{\to} 4 \to 3 & && \to 8 && \to 9 \overset{J}{\to} \text{ ,}
\end{alignat}
respectively. This procedure ensures that there are only finitely many distinct paths $\hat{\gamma}$ and, in particular, that the maximal and minimal values of $\Delta s[\hat{\gamma}]$ are finite. 

Thus, each trimmed path $\hat{\gamma}$ corresponds to a topologically distinct pathway from $I$ to $J$, so that the inequalities \eqref{eq:sec3:aff_result} combine topological and thermodynamic aspects in a similar way as bounds on cycle affinities do, to which they reduce when specifying $I=J$. These aspects are made even more explicit when reformulating the result \eqref{eq:sec3:aff_result} as a statement about the existence of particular microscopic paths in the hidden network. First, a microscopic path $\gamma_+$ from $I$ to $J$ that does not contain any loops or other observed transitions must exist and satisfies the condition
\begin{equation}
    \Delta s[\gamma_+] \geq \sup_t \spath{IJ}{t}.
    \label{eq:sec3:aff_final_var1}
\end{equation}
In other words, not only are there microscopic paths whose entropy production exceeds the value $\sup_t \spath{IJ}{t}$, but such paths can even be found without completing any hidden cycles. A similar existence statement can be made for a path $\gamma_-$ from $I$ to $J$, which satisfies
\begin{equation}
    \Delta s[\gamma_-] \leq \inf_t \spath{IJ}{t}
    \label{eq:sec3:aff_final_var2}
\end{equation}
and again contains neither loops nor any other visible transitions apart from $I$ and $J$.

Before providing an illustration for these results, we briefly discuss their relation to similar-looking bounds published earlier. First, Ref. \cite{degu24} interprets the same quantity $\spath{IJ}{t}$ as a coarse-grained entropy production $\Delta S_{I \to J}(t)$, which satisfies an inequality of the form $\Delta S_{I \to J}(t) \leq \braket{\Delta s[\gamma]|I \overset{t}{\to} J}$ for each value of $t$, where the average is taken over all microscopic paths $\gamma$ that are consistent with the coarse-grained data, i.e., all paths $\gamma$ in the hidden network that start with a transition $I$ followed by a transition $J$ after time $t$. Second, the sequence of inequalities \eqref{eq:sec3:aff_result} is stronger than a conceptually related bound on affinities of hidden cycles also published in Ref. \cite{degu24}, which can be recovered from
\begin{equation}
     \sup_t \spath{IJ}{t} - \inf_t \spath{IJ}{t} \leq \max_{\gamma \in \Gamma} \Delta s[\hat{\gamma}] - \min_{\gamma \in \Gamma} \Delta s[\hat{\gamma}]
     \label{eq:sec3:comparison_prr}
,\end{equation}
a direct consequence of \eqref{eq:sec3:aff_result}. The corresponding result in Ref. \cite{degu24} follows by noting that the existence of paths that realize the maximal and minimal value of $\Delta s$ in the previous equation implies the existence of a circular path with entropy production equal to the right-hand side of Eq. \eqref{eq:sec3:comparison_prr}. If this value is nonzero, we can conclude the existence of a hidden cycle with nonvanishing entropy production.

Third, for $I = J$ the involved trimmed paths $\hat{\gamma}$ start and end with the same transition and therefore correspond to a cycle. In this case, the bounds \eqref{eq:sec3:aff_result} reduce to results introduced in Ref. \cite{vdm22} as affinity estimators for cycles containing the transition $I$ (but no other observed transition). We note that if one is interested in cycle affinities in particular, it is in principle possible to disregard particular observed transitions or include only particular sequences of transitions in the waiting-time distributions to obtain different bounds on cycle affinities. For example, if there are two observed transitions $I$ and $J$ (and their corresponding reverse), we can disregard $J$ to obtain affinity bounds on the set of cycles that contain $I$ irrespective of whether $J$ is included or not. It is also possible to extract waiting-time distributions of, say, the form $\Psi_{I \to I}(t|\text{ contains } J)$. In this case, we could repeat the steps above to find bounds on cycles that contain both transitions $I$ and $J$. The result presented here, however, enables us to treat the paths from $I$ to $J$ and from $J$ to $I$ individually, so that applying the bounds \eqref{eq:sec3:aff_final_var1} or \eqref{eq:sec3:aff_final_var2} will yield more resolved and therefore superior estimates.

\subsection{Numerical illustration}

We illustrate our findings for a network with the topology sketched in \Cref{fig:PBEEexample} a), which has visible transitions $I_\pm$, $J_\pm$ and transition rates as given in \Cref{app:MPSeggpathepr}. We select the transitions $I = I_+ = 7 \to 4$ and $J = J_+ = 9 \to 10$ and depict $\spath{I_+J_+}{t}$ as a function of the waiting time $t$ between the two transitions in \Cref{fig:PBEEexample} b). According to the results \eqref{eq:sec3:aff_final_var1} and \eqref{eq:sec3:aff_final_var2}, there must be one path each from state $4$ to state $9$, whose microscopic entropy production is bounded by $\sup_t \spath{IJ}{t}$ from below and $\inf_t \spath{IJ}{t}$ from above, respectively.

We assess the quality of these bounds by defining an appropriate quality factor $Q$, which satisfies $0 \leq Q \leq 1$ and approaches the upper bound if the corresponding estimate is tight. Additionally, it is sensible to define the quality factor in such a way that $Q$ is invariant under constant shifts of all involved quantities. This is possible because whenever there is a unique shortest path $\gamma_0$ from $I$ to $J$, we have
\begin{equation}
\lim_{t \to 0} \spath{IJ}{t} = \Delta s_{\gamma_0}
\label{eq:sec3:shorttime}
.\end{equation}
Intuitively, this identity states that as $t \to 0$ the probability that the hidden intermediate path between $I$ and $J$ is $\gamma_0$ approaches one. This reasoning can be made more rigorous by expanding $\spath{IJ}{t}$ in powers of $t$ employing similar techniques as in \Cref{sec:shortestpathsL} or Ref. \cite{vdm22}. Thus, we can define 
\begin{equation}
Q = \frac{\sup_t \vert\spath{I_+J_+}{t} - \Delta s_{(4389)}\vert}{\max_{i\in\lbrace 1,2,3 \rbrace}\vert\Delta s_{\gamma_i} - \Delta s_{(4389)}\vert} \label{eq:defQ}
\end{equation}
in the set-up of \Cref{fig:PBEEexample}, where \Cref{eq:sec3:shorttime} ensures that the numerator is operationally accessible. We note that $Q \leq 1$ follows from either \Cref{eq:sec3:aff_final_var1} or \Cref{eq:sec3:aff_final_var2}, depending on whether $\sup_t \spath{IJ}{t}$ or $\inf_t \spath{IJ}{t}$ has a greater distance to $\Delta s_{(4389)}$. In many cases including the example depicted in  \Cref{fig:PBEEexample} b), the other of the two values coincides with $\lim_{t \to 0} \spath{IJ}{t}$, so that the corresponding bound is trivially satisfied.

The results of a numerical study to investigate the performance of $Q$ are depicted in \Cref{fig:PBEEexample} c). The transition rates are randomly drawn from the distributions specified in \Cref{app:MPSeggpathepr}. As can be seen from the blue curve in \Cref{fig:PBEEexample} c), the average value of $Q$ varies between $Q \simeq 0.3$ and $Q \simeq 0.6$, with on average tighter estimates when $\Delta s_{(4389)}$ is large.

\section{Shortest paths between visible transitions}\label{sec:shortestpathsL}

The waiting-time distributions $\wtdPsi{I}{J}{t}$ defined in \Cref{eq:defwtdPsi} for transitions $I=(ij),J=(kl)$ are nonzero
for all times $t >0$ if states $j$ and $k$ are connected in the hidden graph. For such transitions $I$ and $J$, we define
the two topological quantities $\NiIJ{1}{IJ}$ and $\NiIJ{2}{IJ}$. These quantities represent the number of hidden
transitions along the shortest and second-shortest paths between $I$ and $J$, which we require to be self-avoiding. Such
self-avoiding paths naturally contain neither loops nor any repeated transition. This property is inherent to the
shortest path. It is a necessary condition that, in particular, excludes a transition followed by its reverse from the
second-shortest path, so that it corresponds to a distinct path in the underlying graph.
Otherwise, the value $\NiIJ{2}{IJ}$ could not exceed $\NiIJ{1}{IJ}+2$, providing no further insight.

The sum of path weights of all trajectories that start in state $i$ at time $t=0$ and end in the neighboring state $j$
at time $t$ satisfies $p(j,t\vert i,t=0) = k_{ij}t+\mathcal{O}(t)$. For a path with $N_1$ transitions between $j$ and
$l$, the corresponding probability is proportional to $t^{N_1}$ in the short-time limit. Longer paths contribute to
terms of higher order in $t$, while sojourn times enter the path weights in exponential factors. Consequently, with a
shortest path of $N_1$ transitions between $i$ and $l$, we obtain $p(l,t\vert j,t=0) \sim t^{N_1}\left(1 + \mathcal{O}(t)\right)$.
Since $\wtdPsi{I}{J}{t}/k_{kl} = p(l,t\vert j,t=0)$, only the shortest path between $I$ and $J$ contributes to
$\wtdPsi{I}{J}{t}$ in the short-time limit $t\to 0$, which leads to the number of hidden transitions
\begin{align}
\NiIJ{1}{IJ} = \lim_{t\to 0}\left(t\dv{t}\ln\wtdPsi{I}{J}{t}\right) \label{eq:NoneIJ}
\end{align}
along the shortest path from $I$ to $J$ \cite{vdm22}.

Next, we focus on the short-time behavior of the quantity $\spath{IJ}{t}$ defined in \eqref{eq:defaIJ},
\begin{align}
\obsNtwo{IJ} \equiv \lim_{t\to 0}\left(t\dv{t}\ln\abs{\spath{IJ}{t} - \spath{IJ}{0}}\right),
\label{eq:preNtwoIJ}
\end{align}
where the first term of definition \eqref{eq:defaIJ} drops out. This quantity $\obsNtwo{IJ}$ is formally
ill-defined for $\spath{IJ}{t}=const.$, in which case we set $\obsNtwo{IJ}\equiv 0$.

For the remainder of this section, we assume that paths from $I$ to $J$ do not contain a part of a cycle with vanishing affinity.
Within this class, the number of hidden transitions $\NiIJ{2}{IJ}$ along the second-shortest path between $I$ and $J$
is given by
\begin{align}
\NiIJ{2}{IJ} = \obsNtwo{IJ} + \NiIJ{1}{IJ}
\label{eq:NtwoIJ}
\end{align}
in the following cases \cite{vdm22}.

First, this holds true for $\obsNtwo{IJ} \geq 2$, which applies, e.g., to the network shown in \Cref{fig:setupExamples}.

Second, for $\obsNtwo{IJ} = 1$ and $\NiIJ{1}{IJ} \geq 1$, \Cref{eq:NtwoIJ} yields the correct length $\NiIJ{2}{IJ}$ or
the shortest path between $I$ and $J$ is degenerate. In the latter case, the degenerate paths form a cycle with nonzero
affinity, such that for $\NiIJ{1}{IJ} \geq 2$ sojourn times in states along the distinct shortest paths lead to a term
in $\spath{IJ}{t}$ that is proportional to $t^{\NiIJ{1}{IJ}+1}$. Since then the same power arises from a second-shortest
path with $\NiIJ{2}{IJ} = \obsNtwo{IJ} + \NiIJ{1}{IJ}$, this analysis cannot distinguish these two different topologies,
as is the case, e.g., for $I_+\to J_+,\ I_-\to J_-$ and $I_+\to J_-$ in \Cref{fig:PBEEexample}.

\begin{figure}
\centering
\includegraphics[scale=1]{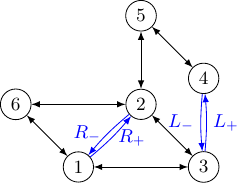}
\caption{
Multicyclic graph of a partially accessible Markov network. Blue, labeled transitions $R_\pm, L_\pm$ are observable,
whereas states and remaining transitions are hidden.
} \label{fig:graphExample}
\end{figure}

Finally, $\obsNtwo{IJ} = 0$, which arises from a time-independent $\spath{IJ}{t}$, implies that there is no second-shortest
self-avoiding path between $I$ and $J$. For example, this is the case in \Cref{fig:graphExample} for $I = L_+$ and $J = R_-$.
Note, however, that this conclusion cannot be drawn if we drop the assumption that paths must not contain parts of cycles with vanishing affinity.

Two additional features will become relevant below. First, $\NiIJ{i}{IJ} = \NiIJ{i}{\tilde{J}\tilde{I}}$ for $i=1,2$ and
$\obsNtwo{IJ} =\obsNtwo{\tilde{J}\tilde{I}}$ because the same path is used in both forward and backward direction. Second,
treating all edges from a set $K$ as hidden can alter the shortest hidden paths between two visible transitions.
For example, ignoring $R_\pm$ transitions in observations of the network depicted in \Cref{fig:graphExample} changes
$\NiIJ{2}{L_\pm L_\pm}$ from $5$ (for the second-shortest path $\overset{L_+}{\to}4\to 5\to 2\to 6\to 1\to 3\overset{L_+}{\to}$)
to $4$ ($\overset{L_+}{\to}4\to 5\to 2\to 1\to 3\overset{L_+}{\to}$). Hence, we define $\MiIJK{1}{IJ}{K}, \obsMtwoK{IJ}{K}$
and $\MiIJK{2}{IJ}{K}$ analogously to \Cref{eq:NoneIJ,eq:preNtwoIJ,eq:NtwoIJ} for observations in which the set $K$
containing the smallest number of visible transitions that decreases the respective number $\NiIJ{1}{IJ}, \obsNtwo{IJ}$
and $\NiIJ{2}{IJ}$ to a specific, different value is considered hidden.

\section{Footprints of clusters: The pair rule}\label{sec:INPwsfclusters}
\begin{figure*}
\centering
\includegraphics[scale=1]{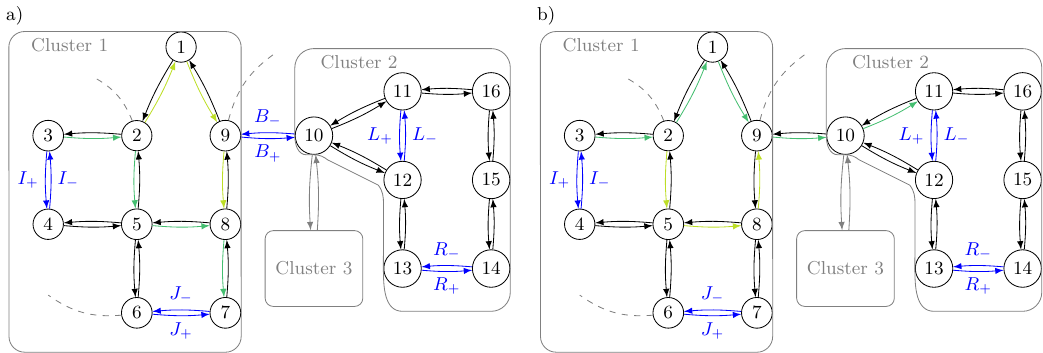}
\caption{Subnetwork with three clusters of states connected by bridges. All states and edges except the blue, labeled
ones are hidden. Dashed, gray lines indicate additional paths to arbitrary subgraphs that are within cluster 1 and
outside the drawn clusters, respectively.
a) Irrespective of whether we can observe edge pair $B_\pm$, which is the bridge between clusters 1 and 2, the shortest
path connecting $I_-$ and $J_-$ is $\overset{I_-}{\to}3\to 2\to 5\to 8\to 7\overset{J_-}{\to}$ shown in teal. The
second-shortest path is $\overset{I_-}{\to} 3\to 2\to 1\to 9\to 8\to 7\overset{J_-}{\to}$ shown in olive green. Their
length differs by $\NiIJ{2}{I_-J_-} - \NiIJ{1}{I_-J_-} = 5 - 4 = 1$ hidden transition.
b) The shortest and the second-shortest path between $I_-$ and $L_+$ consist of teal as well as of olive green edges,
respectively. Both contain the bridge between clusters 1 and 2, and the shortest partial path within cluster 2, while
their partial paths (between states 2 and 9) in cluster 1 are distinct.
} \label{fig:INPclusters}
\end{figure*}

In this section, we assume a network of states that can be characterized as a tree of clusters, see, e.g.,
\Cref{fig:INPclusters}\,a). A cluster consists of states connected by paths. Two clusters are connected by a bridge,
which means they are unconnected after removing the bridge. A genuine bridge, consisting of a pair of edges $B_\pm$,
appears in \Cref{fig:INPclusters}\,a). Within this paper, a bridge could also be larger and consist of a self-avoiding
path of $n$ links. A degenerate bridge is a single state. The tree-like topology on the level of clusters means that
the network obtained by treating all clusters as states does not contain any cycles.

A visible bridge link like $B_\pm$ in \Cref{fig:INPclusters}\,a) allows us to identify clusters
in a partially visible Markov network. Such links $\pm$ are not a part of any cycle within the graph
meaning that both waiting-time distributions $\wtdPsi{+}{+}{t}$ and $\wtdPsi{-}{-}{t}$ vanish identically. While this
condition uniquely characterizes such situations, identifying clusters without having access to visible bridge links
is more involved as we will discuss now.

In preparation for this more general case of inference, we first derive a graph-theoretical fact. For any pair of visible transitions $J_\pm$ and $L_\pm$, consider in each of the four lines
\begin{subequations}
\begin{align}
    &\NiIJ{2}{J_+L_s} - \NiIJ{1}{J_+L_s}, \label{eq:pairi}\\
    &\NiIJ{2}{J_-L_s} - \NiIJ{1}{J_-L_s}, \\
    &\NiIJ{2}{J_sL_+} - \NiIJ{1}{J_sL_+}
\end{align}
and
\begin{align}
    \NiIJ{2}{J_sL_-} - \NiIJ{1}{J_sL_-}\label{eq:pairiv}
\end{align}
\end{subequations}
these differences for $s=+$ and $s=-$, each of which we will call a pair of differences in the following.
If in none of these four pairs the two differences are equal, we can be sure that the two links belong to the same cluster.
Likewise, if the two links belong to different clusters, at least in one pair the two differences have to be equal.
Before we prove this pair rule, as we dub it, we illustrate this criterion with the graph shown in \Cref{fig:INPclusters}.

In this network, we assume $B_+$ and $B_-$ to be hidden in the following. First, consider the edges $I_\pm$ and $J_\pm$.
The shortest and second-shortest path connecting $I_-$ and $J_-$ is highlighted in \Cref{fig:INPclusters}\,a) in teal
and olive green, respectively. We thus get the crucial difference
\begin{align}
    \NiIJ{2}{I_-J_-} - \NiIJ{1}{I_-J_-} &= 5 - 4 = 1.
\end{align}
Similarly, we get
\begin{align}
    \NiIJ{2}{I_-J_+} - \NiIJ{1}{I_-J_+} &= 6 - 3 = 3,
\end{align}
noting that the two differences in this pair are distinct.
Fixing $I_+$ as the first transition, we get
\begin{align}
    \NiIJ{2}{I_+J_-} - \NiIJ{1}{I_+J_-} = 6 - 3 = 3. \label{eq:inpBSP1pm}
\end{align}
For the sequence $I_+\to J_+$, the second-shortest path does not exist. In this case,
we assign the value zero to the difference
\begin{align}
    \NiIJ{2}{I_+J_+} - \NiIJ{1}{I_+J_+} \equiv 0.\label{eq:inpBSP1pp}
\end{align}
Again, we see that the differences in \Cref{eq:inpBSP1pm,eq:inpBSP1pp} are distinct.
Likewise, if we keep the second visible tranisiton fixed and change the first one, i.e.,
$\NiIJ{2}{I_sJ_+} - \NiIJ{1}{I_sJ_+}$ and $\NiIJ{2}{I_sJ_-} - \NiIJ{1}{I_sJ_-}$ for $s=\pm$, we find that in none of
the resulting two pairs with fixed second transition do the differences match. Since $I$ and $J$ are in the same
cluster, the differences in none of the four pairs match, which is in agreement with the pair rule stated above.

Alternatively, if we select the edge pairs $I_\pm$ and $L_\pm$, which are in different clusters, we get for the first pair
\begin{align}
    \NiIJ{2}{I_-L_-} - \NiIJ{1}{I_-L_-} &= 6 - 5 = 1, \notag \\
    \NiIJ{2}{I_-L_+} - \NiIJ{1}{I_-L_+} &= 6 - 5 = 1.
\end{align}
These two differences are the same, since both shortest paths contain the sequence $3\to 2\to 1\to 9$ in Cluster 1, and both second-shortest paths differ from the shortest path only within Cluster 1 by containing the sequence $3\to 2\to 5\to 8\to 9$ instead.
We refer to such parts of shortest paths that lie within one cluster as the shortest partial path and adopt an analogous definition for the second-shortest partial path. The remaining parts of the shortest paths, i.e., the bridge and the partial paths in Cluster 2, corresponding to
each difference are identical and thus do not contribute to it, as shown in \Cref{fig:INPclusters}\,b). We will now show that such an equality between
differences in one of the pairs \eqref{eq:pairi} to \eqref{eq:pairiv} has to occur whenever the chosen pair of visible transitions is in different clusters.

In a general network containing clusters, the second-shortest path between visible edges $J_j,L_l$ with $j,l\in\lbrace +,-\rbrace$ located in different
clusters differs from their shortest path within one of the clusters by the self-avoiding path between bridge and visible
transition. Specifically, it contains the second-shortest partial path, which differs from the corresponding
shortest partial path by the fewest hidden edges. The number of these hidden edges equals the difference
$\NiIJ{2}{J_jL_l} - \NiIJ{1}{J_jL_l}$. Since there are four shortest paths between $J_\pm$ and $L_\pm$ in total, and one
of the differences in partial paths is the smallest, this minimal difference equals
two of the four differences $\NiIJ{2}{J_jL_l} - \NiIJ{1}{J_jL_l}$, thus forming a pair.
Additionally, the transition that remains fixed within the pair indicates the cluster in which the smallest
difference in partial paths between the two visible transitions lies.
Using similar reasoning, if there are two second-smallest differences in partial paths or both second-shortest
partial paths lie in the same cluster, two pairs exist. An example is the combination $I_\pm,L_\pm$
with hidden $B_\pm$ in \Cref{fig:INPclusters}\,b), where the second-shortest partial paths are in Cluster 1.
Other relations between the values of the differences in partial paths lead to a triplet or to two equal pairs.

Consequently, the pair rule yields a sufficient criterion to rule out that two links are in
different clusters. Conversely, a necessary condition for two links to be in different clusters is the existence of
equal differences in at least one such pair of the pair rule. This pair rule thus provides insight into the global structure of a graph.

\section{Reconstructing a minimal graph}\label{sec:constrMinGraph}

The ability to infer topological information about a Markov network based on waiting times
between visible transitions provokes the question of how to reconstruct a minimal graph that is compatible with the
observations.
But what is a minimal graph of such a network that we might want to construct?
We define the largest subgraphs an observer can reconstruct with the tools of \Cref{sec:shortestpathsL,sec:INPwsfclusters}
(up to relabeling states) as a realization of the minimal graph. If this realization is unique it is the minimal graph.

There are three motifs that can be added to a minimal graph without getting into conflict with observations based on the topological information our methods yield.
First, one can always connect otherwise unconnected subgraphs that contain no visible edges to the graph via a hidden bridge as defined in \Cref{sec:INPwsfclusters}. Second, one can always split
a hidden edge into multiple hidden channels between two neighboring states. Third, one can always add paths leading to cycles
with vanishing affinity, unless they introduce a new shortest path that decreases some of the $\NiIJ{1}{\dots}$.
In the following we will focus on the realizations of the minimal graph without such additional elements.
Yet, there may be restrictions or indications that additional extensions are required, the graphical representation of which are then part of the minimal graph.
In summary, the topology of the inferred graph will match the observed data, while in general the full underlying graph may remain
unknown.

Building on operationally accessible knowledge about
shortest paths between edge pairs, in particular the quantities $\NiIJ{1}{\dots}$ and $\obsNtwo{\dots}$ as discussed in
\Cref{sec:shortestpathsL}, we present strategies and tools to reconstruct a minimal graph or realizations thereof
for general, partially accessible Markov networks. First, we discuss the case of one visible pair of edges in
\Cref{sec:CMINGoneEpair}. Second, we illustrate two examples with multiple visible edge pairs, one with and one without
distinct clusters, in \Cref{sec:RGG_cycle_based_app,sec:RGGpba}, respectively. Finally, we present a general algorithm
in \Cref{sec:RGGalgoflow}.

In all cases, we assume the observer has already determined the smallest numbers of hidden transitions between any
two visible ones as discussed in \Cref{sec:shortestpathsL}. Moreover, we again note that hidden paths with the same entropy production as the shortest path
between two visible transitions are not part of a minimal graph, as they lead to a cycle with vanishing affinity. Therefore, we can use \Cref{eq:preNtwoIJ,eq:NtwoIJ} to
determine the hidden length of the second-shortest paths between two visible transitions $I,J$ in the minimal graph,
relying on the results except for the case of $\obsNtwo{IJ}=1$ with $\NiIJ{1}{IJ} \geq 2$.

\subsection{One visible pair of edges}\label{sec:CMINGoneEpair}
\begin{figure}
\centering
\includegraphics[scale=1]{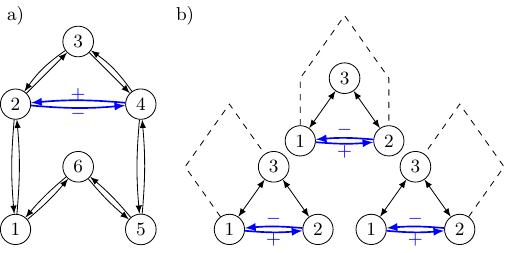}
\caption{A graph and realizations of its minimal graph. Except for the labeled visible edges (blue), all states and edges are
hidden. a) Multicyclic graph with $\NiIJ{1}{++}=2$ and $\NiIJ{2}{++}=4$.
b) The three realizations of a minimal graph for the graph in a). The realizations consist of a three-state unicycle
with labeled visible edges in blue, double-headed arrows as hidden edge pairs and a dashed extension with vertices as
states that potentially exist in the true graph.
} \label{fig:graphExampleExtensions}
\end{figure}

For observations of a single pair of edges, like $+$ and $-$ in \Cref{fig:graphExampleExtensions}\,a), we first note that $\NiIJ{1}{++} > 1$, because otherwise a cycle containing 
only the two states connected by a visible and a hidden link exists. In such a case with visible and hidden channels between states our results can be applied in a similar way, but we will not investigate this case explicitly here.

Proceeding with the case that $\NiIJ{1}{++} > 1$, drawing the shortest cycle of the pair $\pm$ completes the graph that can be reconstructed with certainty.
Since there is no unique way to append the second-shortest cycle of $\pm$, each version of the graph that includes a realization of it is a possible extended minimal graph. Assume, for example,
a network as shown in \Cref{fig:graphExampleExtensions}\,a) where $\NiIJ{1}{++} = 2$ and $\NiIJ{2}{++} = 4$. \Cref{fig:graphExampleExtensions}\,b) displays the three possible realizations of the true graph.

This introductory example demonstrates that a minimal graph need not exist. Instead, we obtain different possible realizations of which one graph is contained in or, in this case, coincides with the underlying graph.

\subsection{Illustrative example I: A cycle-based approach for multiple observed edges}\label{sec:RGG_cycle_based_app}
\begin{figure*}
\centering
\includegraphics[scale=1]{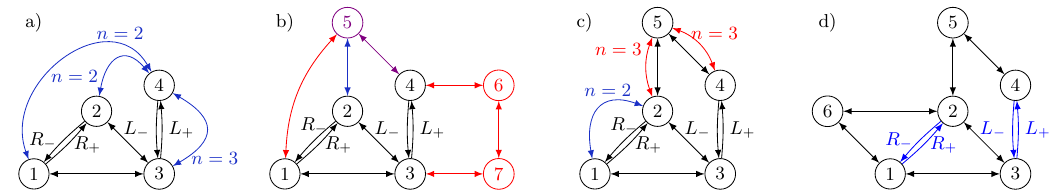}
\caption{Construction of a minimal graph. The procedure relies on the inferred number of hidden transitions
along shortest paths between visible edges as given in \Cref{tab:CBAinferMinGBSPoneTABN}. Visible edge pairs are
labeled, while hidden links are shown as double-headed arrows.
a) The three ways to extend the shortest cycle containing $R_\pm$ and its connection to $L_\pm$ with the shortest cycle
that includes the latter edge pair. Each extension and its number $n$ of required hidden transitions to achieve
$\NiIJ{1}{L_+L_+}=3$ are highlighted in blue.
b) Concrete realizations of extensions in a). The only allowed path is displayed in blue and violet while red
realizations fail to respect $\obsNtwo{L_+R_-}=0$ or $\NiIJ{1}{L_+R_+}=4$.
c) The three ways to extend the unique, compatible graph from b) with the second-shortest cycle containing $L_\pm$ and
$\NiIJ{2}{L_+L_+}=\obsNtwo{L_+L_+}+\NiIJ{1}{L_+L_+}=5$ hidden links that have no similarity with rejected realizations in b). Only the blue path of
$n=2$ hidden transitions is compatible with the values listed in \Cref{tab:CBAinferMinGBSPoneTABN}.
d) Final minimal graph with visible edges in blue. This graph follows after rejecting red extensions in c) and is
the same as the original graph shown in \Cref{fig:graphExample}.} \label{fig:rggCBAexample}
\end{figure*}

In this example, we reconstruct the minimal graph of the Markov network shown in
\Cref{fig:graphExample} by iteratively adding paths and cycles to an initial cycle. We assume the perspective of an observer who has
listed the inferred numbers of hidden transitions along the two shortest paths between visible transitions in
\Cref{tab:CBAinferMinGBSPoneTABN}. Based on this information, the reconstruction proceeds as
follows in three steps, illustrated in \Cref{fig:rggCBAexample}.

\begin{table}
\centering
\caption{Inferred quantities associated with the number of hidden transitions along shortest paths between visible
edges in the network in \Cref{fig:graphExample}. Each number $\NiIJ{1}{\dots},\obsNtwo{\dots} +
\NiIJ{1}{\dots}$ is preceded by one of its two combinations of visible edges like $L_+R_-$, as, e.g., $\NiIJ{1}{L_+R_-} =
\NiIJ{1}{R_+L_-}$ because both associated paths use edges of the same hidden links.}
\label{tab:CBAinferMinGBSPoneTABN}
\begin{tabular}{|c|c|c|c|c|c|c|c|}
\hline
\multicolumn{2}{|c|}{$\NiIJ{1}{\dots}$} & \multicolumn{2}{|c|}{$\NiIJ{1}{\dots}$} &
\multicolumn{2}{|c|}{$\obsNtwo{\dots}+\NiIJ{1}{\dots}$} & \multicolumn{2}{|c|}{$\obsNtwo{\dots}+\NiIJ{1}{\dots}$} \\ \hline
$L_+L_+$ & 3& $\MiIJKedges{L_+L_+}{R_\pm}$ & 3 & $L_+L_+$ & 5& $\MiIJKedges{L_+L_+}{R_\pm}$ & 4 \\
$R_+R_+$ & 2& $\MiIJKedges{R_+R_+}{L_\pm}$ & 2 & $R_+R_+$ & 3& $\MiIJKedges{R_+R_+}{L_\pm}$ & 3 \\
$L_+R_+$ & 4& & & $L_+R_+$ & 5& &  \\
$L_+R_-$ & 2& & & $L_+R_-$ & 2& &  \\
$L_-R_+$ & 1& & & $L_-R_+$ & 3& &  \\
$L_-R_-$ & 1& & & $L_-R_-$ & 3& &  \\ \hline
\end{tabular}
\end{table}

\begin{enumerate}[label=Step \arabic*:\phantom{3.}, ref=\arabic*, wide,series=cbaAppOne]
\item \label{seq:CBA_a1_step12ex} The shortest cycle inferable from \Cref{tab:CBAinferMinGBSPoneTABN} arises from the
    sequence $R_+\to R_+$ with $\NiIJ{1}{R_+R_+} = 2$. It includes $R_\pm$ and two hidden transitions. The state
    between these hidden transitions connects this cycle to the visible edge pair $L_\pm$ as a direct consequence of
    $\NiIJ{1}{L_-R_\pm} = 1$. Thus, we draw this cycle and both visible edge pairs in \Cref{fig:rggCBAexample}\,a),
    where we also indicate all possible ways to add the shortest cycle that contains $L_\pm$.
\item \label{seq:CBA_a1_NEWstep2ex} We need to connect states 2 and 4 via two hidden edge pairs due to
    $\NiIJ{1}{L_+R_-} = 2$. This path, $2\to 5\to 4$, creates a fitting shortest cycle containing $L_\pm$. The two
    alternative ways to draw a shortest cycle of $L_\pm$ fail to preserve $\obsNtwo{L_+R_-} = 0$ as shown in \Cref{fig:rggCBAexample}\,b).
    Moreover, introducing a path of two hidden transitions between states 1 and 4 violates $\NiIJ{1}{L_+R_+} = 4$. After
    rejecting these incompatible realizations, we proceed with the graph displayed in \Cref{fig:rggCBAexample}\,c).
\item \label{seq:CBA_a1_NEWstep3ex} The constructed graph in \Cref{fig:rggCBAexample}\,c) now agrees with all values for
    $\NiIJ{1}{\dots}$ listed in \Cref{tab:CBAinferMinGBSPoneTABN} but at this point lacks appropriate second-shortest paths,
    in particular the second-shortest cycle containing $L_\pm$. This cycle has to comprise $\obsNtwo{L_+L_+} +
    \NiIJ{1}{L_+L_+} = \NiIJ{2}{L_+L_+} = 5$ hidden transitions, for which there are a priori various realizations.
    \Cref{fig:rggCBAexample}\,c) indicates the realizations that we cannot rule out with the previous considerations.
    Since extensions along the path between states 2 and 4 violate $\obsNtwo{L_+R_-} = 0$, a path of
    two hidden transitions between states 1 and 2 is the only realization that is consistent with all data given in
    \Cref{tab:CBAinferMinGBSPoneTABN}. We thus find that $\NiIJ{2}{L_+R_+} = \NiIJ{1}{L_+R_+} = 4$ holds, since
    the corresponding shortest and second-shortest path contain the same number of hidden transitions. This is different
    to what \Cref{tab:CBAinferMinGBSPoneTABN} suggests, i.e., different to $\obsNtwo{L_+R_+} + \NiIJ{1}{L_+R_+} = 1 +
    \NiIJ{1}{L_+R_+} = \NiIJ{2}{L_+R_+}$. Analogously, we find that $\NiIJ{2}{R_+R_+} = \MiIJK{2}{R_+R_+}{L_\pm} =
    \NiIJ{1}{R_+R_+} = 2$ holds. Finally, \Cref{fig:rggCBAexample}\,d) displays the reconstructed minimal graph,
    which matches the original graph from \Cref{fig:graphExample}.
\end{enumerate}
Thus, in this case, the network with six states and eight edge pairs in total can be reconstructed fully based on the
observation of waiting times between combinations of the four transitions that are part of the two visible links.

For more complicated networks that contain clusters and bridges, the steps within this approach often result in multiple
putative graphs. Each of these graphs requires further extension within the method. In such cases, the method branches.
When it comes to connecting potential clusters, we extend and modify this method by focusing on paths rather than cycles.
This adjustment makes the branching behavior more tractable, as will be illustrated in the following example.

\subsection{Illustrative example II: A path-based approach for multiple observed edges}\label{sec:RGGpba}
\begin{figure*}
\centering
\includegraphics[scale=1]{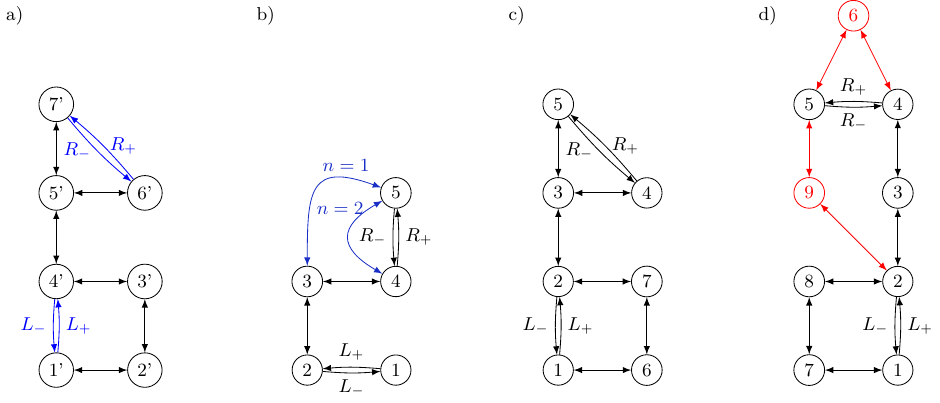}
\caption{Construction of minimal graphs for a graph with two clusters.
To stress the unknown character, the states of the underlying graph are labeled with a prime, whereas in the
reconstruction the states are labeled with consecutive numbers.
a) Graph with visible transitions $R_\pm,L_\pm$ that leads to the inferred data listed in \Cref{tab:PBAinferMinGBSPoneTABN}.
Using this data, we construct minimal graphs in b) to d), which demonstrates the branching behavior of this approach
leading to multiple realizations for a graph.
b) A shortest hidden path between $L_\pm$ and $R_\pm$ with $\NiIJ{1}{L_+R_+} = 2$. The blue labeled arrows indicate the
two possible ways to draw the shortest cycle that contains $R_\pm$ and $\NiIJ{1}{R_+R_+} = 2$, with $n=1,2$ additional,
hidden transitions.
c) Realization of a minimal graph with $n = 1$ from b) in which the shortest cycle containing $L_\pm$ with
$\NiIJ{1}{L_+L_+} = 3$ must not include
states 4,5 and 3, since a direct link between states 1 and $4, 5$ as well as between 6 and 3 would contradict
$\NiIJ{1}{L_-R_+} = 5$.
d) Realization of a minimal graph with $n = 2$ in b) with shortest cycle containing $R_\pm$ leading to
$\NiIJ{1}{R_+R_+} = 2$ and shortest hidden path between $L_+$ and $R_-$ leading to $\NiIJ{1}{L_+R_-} = 2$ in red.
The red paths form a second-shortest cycle including $R_+$ and a second-shortest path for the four
sequences $L_\pm\to R_\pm$ that are in conflict with the corresponding $\obsNtwo{\dots} = 0$. Alternatively, the affinity-based
criterion discussed in \Cref{sec:RMGFTepra} helps to rule out this realization since all $\spath{\dots}{t}$ are
time-independent such that the graph in c) remains as the unique minimal graph that is the same as the graph in a) after
relabeling states.
} \label{fig:rggPBAexample}
\end{figure*}
\begin{table}
\centering
\caption{Inferred numbers of hidden transitions along shortest paths between two visible transitions in a
graph as shown in \Cref{fig:rggPBAexample}\,a). The combination of visible edges precedes its values $\NiIJ{1}{\dots}$.
}
\label{tab:PBAinferMinGBSPoneTABN}
\begin{tabular}{|c|c|c|c|}
\hline
\multicolumn{2}{|c|}{$\NiIJ{1}{\dots}$} & \multicolumn{2}{|c|}{$\NiIJ{1}{\dots}$} \\ \hline
$L_+L_+$ & 3& $L_+R_+$ & 2 \\
$\MiIJKedges{L_+L_+}{R_\pm}$ & 3 & $L_+R_-$ & 2 \\
$R_+R_+$ & 2& $L_-R_+$ & 5 \\
$\MiIJKedges{R_+R_+}{L_\pm}$ & 2 & $L_-R_-$ & 5 \\ \hline
\end{tabular}
\end{table}

In our second example, we focus on paths between distinct visible edge pairs, which in this case are key to connect visible
edge pairs from different clusters. This example also elucidates the type of branching behavior that results in different
compatible graphs. We consider the graph shown in \Cref{fig:rggPBAexample}\,a), which contains two clusters. The information
on shortest paths that an observer can infer is given in \Cref{tab:PBAinferMinGBSPoneTABN}. For this network, the
observer finds that $\obsNtwo{\dots} = 0$ for all sequences of transitions. Thus, the minimal graph will contain no
second-shortest paths.

\begin{enumerate}[label=Step \arabic*:\phantom{3.}, ref=\arabic*, wide,series=pbaAppOneEx]
\item \label{itm:PBA_stepaaex}
    The sequence $L_+\to R_+$ has one of the two equally shortest paths with $\NiIJ{1}{L_+R_+} = 2 = \NiIJ{1}{L_+R_-}$
    hidden links between the two links $L_\pm,R_\pm$, as given in the second column of \Cref{tab:PBAinferMinGBSPoneTABN}.
    We therefore start drawing the graph with this sequence, as depicted in \Cref{fig:rggPBAexample}\,b).
\item \label{itm:PBA_stepaex} The shortest cycle including $R_\pm$ and $\NiIJ{1}{R_+R_+} = 2$ hidden links can be
    added in the two ways indicated in \Cref{fig:rggPBAexample}\,b). We start with the realization featuring $n = 1$
    hidden link between states 3 and 5. Adding the shortest cycle that contains $L_\pm$ and $\NiIJ{1}{L_+L_+} = 3$
    hidden links, as displayed in \Cref{fig:rggPBAexample}\,c), completes an allowed minimal graph.
    In particular, this graph includes neither shorter self-avoiding paths than given by the values in
    \Cref{tab:PBAinferMinGBSPoneTABN} nor second-shortest paths. Other ways to draw this cycle
    contradict $\NiIJ{1}{L_-R_+}= 5$. Therefore, the only resulting graph matches the original one in
    \Cref{fig:rggPBAexample}\,a) up to state relabeling.
    
    A putative further realization uses the $n = 2$ option from \Cref{fig:rggPBAexample}\,b). This realization again
    requires connecting states 1 and 2 as shown in \Cref{fig:rggPBAexample}\,c) to the unique shortest cycle containing
    $L_\pm$, and ensuring that second-shortest cycles that contradict \Cref{tab:PBAinferMinGBSPoneTABN} are not introduced.
    The value $\NiIJ{1}{L_+R_-} = 2$ requires a
    separate path of two hidden transitions between $L_+$ and $R_-$ like in \Cref{fig:rggPBAexample}\,d), but this
    introduces an unwanted second-shortest cycle containing $R_\pm$. Drawing this path would also conflict
    with $\obsNtwo{\dots} = 0$ for the four sequences $L_\pm\to R_\pm$ because they have no second-shortest paths.
    Thus, this realization is incompatible with the observations and must be discarded.
\end{enumerate}

The approach described above is not the only way to obtain the graph that yields the desired values of $\NiIJ{1}{\dots}$
as given in \Cref{tab:PBAinferMinGBSPoneTABN} and $\obsNtwo{\dots} = 0$. We could alternatively use a criterion discussed
in detail in \Cref{sec:RMGFTepra}, which allows us to restrict the set of compatible graphs by using methods similar to
those described in \Cref{sec:INPpbeEstimation}. The reasoning is based on inferring cycle affinities and entropy production
along microscopic paths and discarding the realizations where these quantities lead to inconsistent results.

In conclusion, we are able to find a unique minimal graph that, in this case, coincides with the original graph after
relabeling states. This example also demonstrates that even if we obtain different possible realizations of the minimal
graph in an intermediate step, it is possible that such realizations have to be discarded at a later stage when they
turn out to be inconsistent.

\begin{figure*}
\centering
\includegraphics[scale=1]{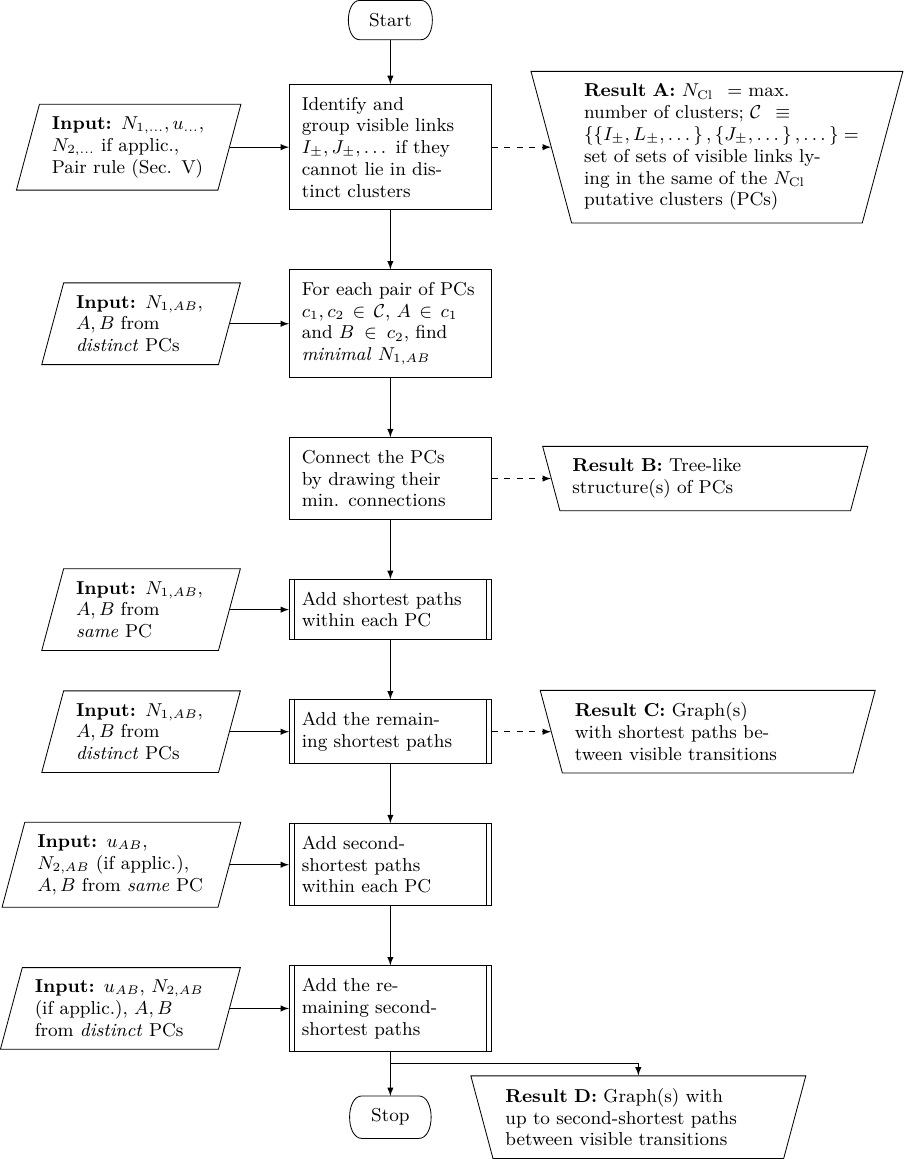}
\caption{%
An algorithm to reconstruct realizations of a minimal graph. The procedure takes the inferred
values of $\NiIJ{1}{\dots},\obsNtwo{\dots}$ and, if applicable, of $\NiIJ{2}{\dots}$ along with the pair rule from
\Cref{sec:INPwsfclusters} as its input. It is possible that the algorithm branches, therefore it may generate multiple putative realizations of the minimal graph in the general case. The outlined procedure has to be followed for every realization individually in this case.
For convenience, we list four intermediate results A to D, resulting in a four-block structure. The first block A yields the maximum number $N_\text{Cl}$ of putative clusters (PCs) and a set $\mathcal{C}$ containing sets of visible links that are in the same PC. 
Starting in the line below result A, we connect the PCs in block B via the paths with the minimal $\NiIJ{1}{\dots}$ between each pair of PCs. The result are realizations of graphs with tree-like topology on the level of PCs. In the last two blocks, we repeatedly call the subroutine defined in flowchart \ref{fig:flowchartsubr}, which is indicated by the box with double vertical lines. Explicitly given inputs define the task described in the
label of the subroutine, which extends putative realizations of a minimal graph with given input. Block C generates shortest paths between all visible transitions in all putative
realizations, while ensuring that they respect all relevant $\NiIJ{1}{\dots}$ and $\obsNtwo{\dots}$. Note that checking
for additional realizations between visible transitions is crucial if two PCs are
merged. Block D includes the analogous procedure for the second-shortest connections.}
\label{fig:flowchart}
\end{figure*}

\subsection{General method to reconstruct realizations of a minimal graph}\label{sec:RGGalgoflow}

In this section, we compile our previous heuristic insights into an algorithm, which is illustrated as a flowchart in Figure \ref{fig:flowchart}.

We can conceptualize the algorithm as comprising two distinct components. The first part encompasses the first three steps in the algorithm until Result B and aims at identifying putative clusters (PCs) based on the pair rule of \Cref{sec:INPwsfclusters}. These clusters are connected before we proceed with the connections within a PC. This approach proved useful for networks that feature bridges or bottlenecks like the one previously discussed in \Cref{sec:RGGpba}. It is worth noting that, since the pair rule only provides a necessary criterion to identify multiple clusters, the structure of PCs may change as additional information is added to the subgraph later in the algorithm. For example, if it turns out that two PCs are connected by not only one bridge, these two PCs have to be combined into a single PC, which also entails checking whether there are additional cross-connections between the previously separate clusters.

\begin{figure}
\centering
\includegraphics[scale=1]{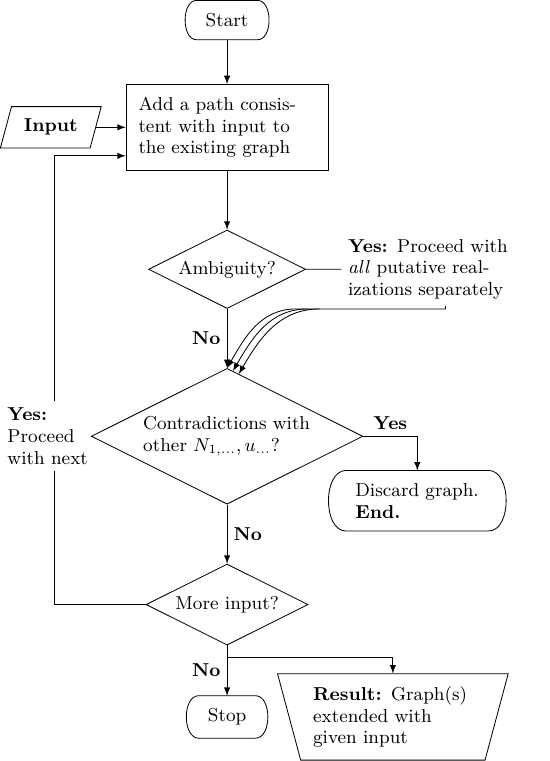}
\caption{%
The subroutine used in Figure \ref{fig:flowchart}. The explicitly given input of the subroutine defines
its task and its label in flowchart \ref{fig:flowchart}, but all $\NiIJ{1}{\dots},\obsNtwo{\dots}$
are implicitly used as input when checking for contradictions. 
Based on the explicitly given input, the subroutine adds
paths to each putative realization. If multiple realizations are possible, we proceed with each putative realization and corresponding graph separately. 
Realizations that do not respect all $\NiIJ{1}{\dots}$ and $\obsNtwo{\dots}$ are discarded. 
The subroutine ultimately returns all resulting putative realizations, now extended by the given input.}
\label{fig:flowchartsubr}
\end{figure}

The second part of the algorithm utilizes this preliminary structure to systematically include information about first the shortest and then the second-shortest connections. This procedure repeatedly calls the subroutine depicted in \Cref{fig:flowchartsubr}. In this subroutine we encounter the peculiar branching behavior that was already present in the previous examples. It is important to note that the algorithm must be continued for each individual putative realization, because it is possible that such branches can be ruled out at some later point in the algorithm, as we encountered in the example of \Cref{sec:RGGpba}. As one would expect already from the introductory example in \Cref{sec:CMINGoneEpair},  the algorithm yields not a single minimal graph but different possible ones in the general case.

A more formal presentation of the algorithm is found as pseudocode in \Cref{sec:RGGalgo}, which also discusses particularly subtle aspects in more detail.

\section{Discussion: Conceptual approaches and their limitations} \label{sec:discussion}

This work focuses on what might be deemed a thermodynamic approach to inference of an underlying graph. The theme common to the methods described here is, apart from assuming that transitions are the elementary observable, the conceptual focus on $\spath{IJ}{t}$. This measure of broken detailed balance that can be given a precise thermodynamic interpretation as a coarse-grained entropy production \cite{degu24}. In this sense, we provide a quantitative extension of previous studies that relate observed broken time-reversal symmetry, a qualitative, binary property, to topological aspects of the underlying hidden layer \cite{bere06, bere19, glad19}.

An important consequence is that the methods presented here are insensitive to a number of operations that do not change nonequilibrium characteristics, which includes the addition of equilibrium cycles, tree-like appendages or even entire clusters if such clusters do not contain a single visible transition and connect to the remaining network only via a bridge. Additionally, a hidden edge could always be split into multiple channels between two states. Due to such shortcomings, one might argue that the thermodynamic approach is inferior to, say, Markov state modeling \cite{bowm14}, because the reconstruction of an underlying model is at best possible in a ``minimal'' sense, i.e., up to in this approach invisible additions.

However, if we are particularly interested in the nonequilibrium aspects of a given model, insensitivity to changes in the nondriven parts of the network may be a desirable rather than a detrimental feature. To illustrate this point, let us consider the heat shock protein Hsp90 as a concrete example. This protein displays nonequilibrium behavior in the presence of ATP and suitable cochaperones and is therefore interesting from a thermodynamic point of view \cite{voll24a}, but the presence of strong memory effects in the observed dynamics render a direct description in terms of a hidden Markov model difficult and potentially even misleading \cite{voll24}. In such a scenario, our thermodynamic tools offer insights into the parts of the underlying model that couple to the external driving. In combination with other constructive methods like hidden Markov modeling, our tools can also be understood as a set of operationally verifiable criteria to check whether a proposed model allows for a consistent description from a thermodynamic point of view.

In the following, we comment on the limitations of our methods in more detail. These remarks serve as possible starting points for improvements and refinements of our methods that may be the subject of future work.

\subsection{Thermodynamic inference of microscopic paths}

In \Cref{sec:INPpbeEstimation} we present a set of inequalities through which we can infer the existence of microscopic paths with dissipation greater or smaller than an operationally accessible threshold. Focusing on properties of the model on the underlying Markovian level, these relations complement a number of results regarding thermodynamic properties of the transition-based model on the coarse-grained level \cite{vdm22, haru22, haru23a, neri23, degu24, degu24b}. 

The main difficulty in applying these results lies in identifying an appropriate underlying Markovian layer to which these results can be applied. From a theoretical point of view, it is necessary that the dynamics and energetics of the underlying Markov model are related as described by the laws of stochastic thermodynamics \cite{seif12}. In practice, such insights require knowledge about the physical coupling and energy transduction between the degrees of freedom that comprise the system and its environment. A promising example where such methods might be useful are molecular motors, where microscopic models that describe their energetics adequately are well-known \cite{juel97, gasp07, zimm12}. In such a case, we might speculate whether additional insights into the energetics (e.g. that cycle affinities are integer multiples of the free energy available through consumption of one ATP molecule) allow even stronger predictions about the microscopic model (e.g. the existence of a microscopic path along which one or several ATP molecules are consumed).

\subsection{Number of hidden states along shortest paths}

Our subsequent, more structural results are based on the observation that the short-time limit of waiting-time distributions contains information about the number of hidden intermediate states as described quantitatively in \Cref{sec:shortestpathsL}. From a practical standpoint, a significant obstacle may be to first acquire sufficient statistics to sample a waiting-time distribution and subsequently extract its short-time limit. Additionally, extracting the number of hidden states in the second-shortest cycle from the quantity $\spath{IJ}{t}$ defined in \Cref{eq:preNtwoIJ} is not always possible. As a nonequilibrium quantity with thermodynamic meaning, $\spath{IJ}{t}$ vanishes in equilibrium and is also otherwise unsensitive to some operations that do not change nonequilibrium properties of the network. More specifically, the apparent number of states along the second-shortest path does not change if one adds additional equilibrium cycles to the network, i.e., if the shortest and second-shortest hidden path produce the same amount of entropy, then the number of hidden states in the second-shortest path cannot be inferred.

\subsection{Detection of clusters}

The detection of clusters through what we have called the pair rule in \Cref{sec:INPwsfclusters} demonstrates that ``local'' inference methods based on shortest paths between transitions can in some cases provide insight into ``global'' properties of the network as well. Violation of the pair rule gives a sufficient criterion to rule out when two pairs of observed transitions lie in different clusters. However, we point out two important limitations of this approach. First, the pair rule is only a necessary criterion that the two links are located in different clusters. Second, in our reasoning we assume that if we treat the clusters as lumped states, the resulting network containing only the clusters does not contain any cycles. The design of more refined methods to detect clusters remains as a challenging open problem for thermodynamic inference, which given the previously discussed general limitations of thermodynamic inference might require a different conceptual approach.

\subsection{Reconstruction of a minimal graph}

The construction of nontrivial minimal graphs relies heavily on access to multiple observed transitions and the short-time limit of their corresponding waiting-time distributions, which are essential for inferring the number of hidden transitions along the shortest connecting paths. Assuming we have information on the number of states in both the shortest and second-shortest paths between pairs of transitions, the methods in \Cref{sec:constrMinGraph} aim at utilizing as many of the inferred quantities as possible in a systematic and consistent way. However, some difficulties inherent to this approach must be addressed.

The perhaps most severe problem is that in the general case the algorithm branches repeatedly as more topological information is included. Thus, one is left with not a single minimal graph but multiple possibilities which might be discarded at some later point in the algorithm. It is generally to be expected that this branching behavior is more severe for more strongly connected graphs, i.e., graphs in which a vertex is connected to more hidden edges. In the general case, there is also no guarantee that the minimal graph is unique so that in the worst case one is left with a large number of potential minimal graphs corresponding to the different surviving branches of the algorithm.

\section{Concluding remarks and outlook} \label{sec:outlook}

This work describes different ways to utilize the observation of transition statistics in a partially accessible Markov network, but by no means exhausts all possibilities, because the statistics contains additional accessible but unused information.
For example, the shape of the curve $\spath{IJ}{t}$ generally is simpler if the number of topologically distinct pathways from $I$ to $J$ is small, to the point that $\spath{IJ}{t}$ is constant in time if there is only one pathway from $I$ to $J$. One might therefore speculate whether more complicated shapes of $\spath{IJ}{t}$ reveal additional topological data through their number of extrema, which could originate either from decaying coherent oscillations or topologically distinct hidden paths between the visible transitions $I,J$. In the latter case, the conjecture aligns with an observation in Ref. \cite{vdm22} for the case $I=J$, providing an example in which each of the aforementioned extrema appears to bound the affinity of a different cycle containing the transition $I$.

From the perspective of graph reconstruction as discussed in \Cref{sec:constrMinGraph}, it is also worth investigating other methods to infer the number of transitions along hidden pathways. This may take the form of a systematic analysis of the short-time limit, which then encodes information about successively longer paths in terms of successively higher order in time. However, such an approach faces not only conceptual limitations but also the practical challenge of extracting such information from experimental data. Nevertheless, additional structural information about path lengths is beneficial in multiple ways. First, we gain the ability to infer a larger portion of the underlying graph. Second, additional information provides additional restrictions on minimal graphs that can reduce the number of allowed realizations of graphs and therefore reduce the complexity of our proposed algorithm. 

We may also speculate about other model classes beyond the case of observing transitions in the steady state. Due to the conceptual focus of our work, we assume systems with constant driving for simplicity, but the thermodynamic formalism based on waiting-time distributions can be transferred to the time-dependent case \cite{maie24, haru24b} and even to a more general class of observables beyond transitions \cite{degu24b}. We therefore expect that our results regarding hidden paths and graphs can be in principle generalized to such settings as well. Non-Markovian observables, such as lumped states or events whose reverse is not accessible, e.g., transitions for which only one direction is observed, comprise two classes of observables that have gained recent interest \cite{gode23, igos25}. Another way to generalize the presented work is to allow for nonexponential dwell times within a state, which is a way to incorporate memory in the system. In this scenario, a starting point for future studies is Ref. \cite{li14}, which makes use of waiting-time distributions to extract the number of hidden transitions along the shortest path between two succeeding visible states.

As a final remark, we emphasize that our techniques for graph reconstruction provide only one possible approach. Thus, our proof of concept provides a basis for further studies to design and investigate improved or even optimal algorithms to construct a minimal graph. The precise characterization of a minimal graph (or set of potential minimal graphs) that at the same time provides a thermodynamically consistent description is a challenging task that will be left for future works.

\section*{Acknowledgments}
We thank Julius Degünther for insightful discussions. J.v.d.M. was partially supported by JSPS KAKENHI (Grant No. 24H00833).

%
%
%
%
%
%
%
%
%
%
%
%
%
%
%
%
%
%
%
\appendix\crefalias{section}{appendix}\crefalias{subsection}{appendix}
\bibliography{references.bib}

\begin{thebibliography}{56}%
\makeatletter
\providecommand \@ifxundefined [1]{%
 \@ifx{#1\undefined}
}%
\providecommand \@ifnum [1]{%
 \ifnum #1\expandafter \@firstoftwo
 \else \expandafter \@secondoftwo
 \fi
}%
\providecommand \@ifx [1]{%
 \ifx #1\expandafter \@firstoftwo
 \else \expandafter \@secondoftwo
 \fi
}%
\providecommand \natexlab [1]{#1}%
\providecommand \enquote  [1]{``#1''}%
\providecommand \bibnamefont  [1]{#1}%
\providecommand \bibfnamefont [1]{#1}%
\providecommand \citenamefont [1]{#1}%
\providecommand \href@noop [0]{\@secondoftwo}%
\providecommand \href [0]{\begingroup \@sanitize@url \@href}%
\providecommand \@href[1]{\@@startlink{#1}\@@href}%
\providecommand \@@href[1]{\endgroup#1\@@endlink}%
\providecommand \@sanitize@url [0]{\catcode `\\12\catcode `\$12\catcode
  `\&12\catcode `\#12\catcode `\^12\catcode `\_12\catcode `\%12\relax}%
\providecommand \@@startlink[1]{}%
\providecommand \@@endlink[0]{}%
\providecommand \url  [0]{\begingroup\@sanitize@url \@url }%
\providecommand \@url [1]{\endgroup\@href {#1}{\urlprefix }}%
\providecommand \urlprefix  [0]{URL }%
\providecommand \Eprint [0]{\href }%
\providecommand \doibase [0]{https://doi.org/}%
\providecommand \selectlanguage [0]{\@gobble}%
\providecommand \bibinfo  [0]{\@secondoftwo}%
\providecommand \bibfield  [0]{\@secondoftwo}%
\providecommand \translation [1]{[#1]}%
\providecommand \BibitemOpen [0]{}%
\providecommand \bibitemStop [0]{}%
\providecommand \bibitemNoStop [0]{.\EOS\space}%
\providecommand \EOS [0]{\spacefactor3000\relax}%
\providecommand \BibitemShut  [1]{\csname bibitem#1\endcsname}%
\let\auto@bib@innerbib\@empty
\bibitem [{\citenamefont {Sekimoto}(2010)}]{seki10}%
  \BibitemOpen
  \bibfield  {author} {\bibinfo {author} {\bibfnamefont {K.}~\bibnamefont
  {Sekimoto}},\ }\href {https://doi.org/10.1007/978-3-642-05411-2} {\emph
  {\bibinfo {title} {Stochastic Energetics}}}\ (\bibinfo  {publisher}
  {Springer},\ \bibinfo {address} {Berlin, Heidelberg},\ \bibinfo {year}
  {2010})\BibitemShut {NoStop}%
\bibitem [{\citenamefont {{J}arzynski}(2011)}]{jarz11}%
  \BibitemOpen
  \bibfield  {author} {\bibinfo {author} {\bibfnamefont {C.}~\bibnamefont
  {{J}arzynski}},\ }\bibfield  {title} {\bibinfo {title} {Equalities and
  inequalities: Irreversibility and the second law of thermodynamics at the
  nanoscale},\ }\href
  {https://doi.org/10.1146/annurev-conmatphys-062910-140506} {\bibfield
  {journal} {\bibinfo  {journal} {Ann. Rev. Cond. Mat. Phys.}\ }\textbf
  {\bibinfo {volume} {2}},\ \bibinfo {pages} {329} (\bibinfo {year}
  {2011})}\BibitemShut {NoStop}%
\bibitem [{\citenamefont {Seifert}(2012)}]{seif12}%
  \BibitemOpen
  \bibfield  {author} {\bibinfo {author} {\bibfnamefont {U.}~\bibnamefont
  {Seifert}},\ }\bibfield  {title} {\bibinfo {title} {Stochastic
  thermodynamics, fluctuation theorems, and molecular machines},\ }\href
  {https://doi.org/10.1088/0034-4885/75/12/126001} {\bibfield  {journal}
  {\bibinfo  {journal} {Rep. Prog. Phys.}\ }\textbf {\bibinfo {volume} {75}},\
  \bibinfo {pages} {126001} (\bibinfo {year} {2012})}\BibitemShut {NoStop}%
\bibitem [{\citenamefont {Peliti}\ and\ \citenamefont
  {Pigolotti}(2021)}]{peli21}%
  \BibitemOpen
  \bibfield  {author} {\bibinfo {author} {\bibfnamefont {L.}~\bibnamefont
  {Peliti}}\ and\ \bibinfo {author} {\bibfnamefont {S.}~\bibnamefont
  {Pigolotti}},\ }\href@noop {} {\emph {\bibinfo {title} {Stochastic
  thermodynamics. An Introduction}}}\ (\bibinfo  {publisher} {Princeton Univ.
  Press},\ \bibinfo {address} {Princeton and {O}xford},\ \bibinfo {year}
  {2021})\BibitemShut {NoStop}%
\bibitem [{\citenamefont {Shiraishi}(2023{\natexlab{a}})}]{shir23}%
  \BibitemOpen
  \bibfield  {author} {\bibinfo {author} {\bibfnamefont {N.}~\bibnamefont
  {Shiraishi}},\ }\href {https://doi.org/10.1007/978-981-19-8186-9} {\emph
  {\bibinfo {title} {An Introduction to Stochastic Thermodynamics}}},\
  Fundamental Theories of Physics\ (\bibinfo  {publisher} {Springer Nature
  Singapore},\ \bibinfo {address} {Singapore},\ \bibinfo {year}
  {2023})\BibitemShut {NoStop}%
\bibitem [{\citenamefont {Seifert}(2019)}]{seif19}%
  \BibitemOpen
  \bibfield  {author} {\bibinfo {author} {\bibfnamefont {U.}~\bibnamefont
  {Seifert}},\ }\bibfield  {title} {\bibinfo {title} {From stochastic
  thermodynamics to thermodynamic inference},\ }\href
  {https://doi.org/10.1146/annurev-conmatphys-031218-013554} {\bibfield
  {journal} {\bibinfo  {journal} {Ann. Rev. Cond. Mat. Phys.}\ }\textbf
  {\bibinfo {volume} {10}},\ \bibinfo {pages} {171} (\bibinfo {year}
  {2019})}\BibitemShut {NoStop}%
\bibitem [{\citenamefont {Barato}\ and\ \citenamefont
  {Seifert}(2015)}]{seif15}%
  \BibitemOpen
  \bibfield  {author} {\bibinfo {author} {\bibfnamefont {A.~C.}\ \bibnamefont
  {Barato}}\ and\ \bibinfo {author} {\bibfnamefont {U.}~\bibnamefont
  {Seifert}},\ }\bibfield  {title} {\bibinfo {title} {Thermodynamic uncertainty
  relation for biomolecular processes},\ }\href
  {https://doi.org/10.1103/PhysRevLett.114.158101} {\bibfield  {journal}
  {\bibinfo  {journal} {Phys. Rev. Lett.}\ }\textbf {\bibinfo {volume} {114}},\
  \bibinfo {pages} {158101} (\bibinfo {year} {2015})}\BibitemShut {NoStop}%
\bibitem [{\citenamefont {Gingrich}\ \emph {et~al.}(2016)\citenamefont
  {Gingrich}, \citenamefont {Horowitz}, \citenamefont {Perunov},\ and\
  \citenamefont {England}}]{ging16}%
  \BibitemOpen
  \bibfield  {author} {\bibinfo {author} {\bibfnamefont {T.~R.}\ \bibnamefont
  {Gingrich}}, \bibinfo {author} {\bibfnamefont {J.~M.}\ \bibnamefont
  {Horowitz}}, \bibinfo {author} {\bibfnamefont {N.}~\bibnamefont {Perunov}},\
  and\ \bibinfo {author} {\bibfnamefont {J.~L.}\ \bibnamefont {England}},\
  }\bibfield  {title} {\bibinfo {title} {Dissipation bounds all steady-state
  current fluctuations},\ }\href
  {https://doi.org/10.1103/PhysRevLett.116.120601} {\bibfield  {journal}
  {\bibinfo  {journal} {Phys. Rev. Lett.}\ }\textbf {\bibinfo {volume} {116}},\
  \bibinfo {pages} {120601} (\bibinfo {year} {2016})}\BibitemShut {NoStop}%
\bibitem [{\citenamefont {Horowitz}\ and\ \citenamefont
  {Gingrich}(2020)}]{horo20}%
  \BibitemOpen
  \bibfield  {author} {\bibinfo {author} {\bibfnamefont {J.~M.}\ \bibnamefont
  {Horowitz}}\ and\ \bibinfo {author} {\bibfnamefont {T.~R.}\ \bibnamefont
  {Gingrich}},\ }\bibfield  {title} {\bibinfo {title} {Thermodynamic
  uncertainty relations constrain non-equilibrium fluctuations},\ }\href
  {https://doi.org/https://doi.org/10.1038/s41567-019-0702-6} {\bibfield
  {journal} {\bibinfo  {journal} {Nat. Phys.}\ }\textbf {\bibinfo {volume}
  {16}},\ \bibinfo {pages} {15} (\bibinfo {year} {2020})}\BibitemShut {NoStop}%
\bibitem [{\citenamefont {Shiraishi}\ \emph {et~al.}(2018)\citenamefont
  {Shiraishi}, \citenamefont {Funo},\ and\ \citenamefont {Saito}}]{shir18}%
  \BibitemOpen
  \bibfield  {author} {\bibinfo {author} {\bibfnamefont {N.}~\bibnamefont
  {Shiraishi}}, \bibinfo {author} {\bibfnamefont {K.}~\bibnamefont {Funo}},\
  and\ \bibinfo {author} {\bibfnamefont {K.}~\bibnamefont {Saito}},\ }\bibfield
   {title} {\bibinfo {title} {Speed limit for classical stochastic processes},\
  }\href {https://doi.org/10.1103/PhysRevLett.121.070601} {\bibfield  {journal}
  {\bibinfo  {journal} {Phys. Rev. Lett.}\ }\textbf {\bibinfo {volume} {121}},\
  \bibinfo {pages} {070601} (\bibinfo {year} {2018})}\BibitemShut {NoStop}%
\bibitem [{\citenamefont {Ito}\ and\ \citenamefont {Dechant}(2020)}]{ito20}%
  \BibitemOpen
  \bibfield  {author} {\bibinfo {author} {\bibfnamefont {S.}~\bibnamefont
  {Ito}}\ and\ \bibinfo {author} {\bibfnamefont {A.}~\bibnamefont {Dechant}},\
  }\bibfield  {title} {\bibinfo {title} {Stochastic {{Time Evolution}},
  {{Information Geometry}}, and the {{Cram{\'e}r-Rao Bound}}},\ }\href
  {https://doi.org/10.1103/PhysRevX.10.021056} {\bibfield  {journal} {\bibinfo
  {journal} {Physical Review X}\ }\textbf {\bibinfo {volume} {10}},\ \bibinfo
  {pages} {021056} (\bibinfo {year} {2020})}\BibitemShut {NoStop}%
\bibitem [{\citenamefont {Dechant}(2022)}]{dech22a}%
  \BibitemOpen
  \bibfield  {author} {\bibinfo {author} {\bibfnamefont {A.}~\bibnamefont
  {Dechant}},\ }\bibfield  {title} {\bibinfo {title} {Minimum entropy
  production, detailed balance and {{Wasserstein}} distance for continuous-time
  {{Markov}} processes},\ }\href {https://doi.org/10.1088/1751-8121/ac4ac0}
  {\bibfield  {journal} {\bibinfo  {journal} {Journal of Physics A:
  Mathematical and Theoretical}\ }\textbf {\bibinfo {volume} {55}},\ \bibinfo
  {pages} {094001} (\bibinfo {year} {2022})}\BibitemShut {NoStop}%
\bibitem [{\citenamefont {Van~Vu}\ and\ \citenamefont {Saito}(2023)}]{vu23a}%
  \BibitemOpen
  \bibfield  {author} {\bibinfo {author} {\bibfnamefont {T.}~\bibnamefont
  {Van~Vu}}\ and\ \bibinfo {author} {\bibfnamefont {K.}~\bibnamefont {Saito}},\
  }\bibfield  {title} {\bibinfo {title} {Thermodynamic {{Unification}} of
  {{Optimal Transport}}: {{Thermodynamic Uncertainty Relation}}, {{Minimum
  Dissipation}}, and {{Thermodynamic Speed Limits}}},\ }\href
  {https://doi.org/10.1103/PhysRevX.13.011013} {\bibfield  {journal} {\bibinfo
  {journal} {Physical Review X}\ }\textbf {\bibinfo {volume} {13}},\ \bibinfo
  {pages} {011013} (\bibinfo {year} {2023})}\BibitemShut {NoStop}%
\bibitem [{\citenamefont {Nagayama}\ \emph {et~al.}(2025)\citenamefont
  {Nagayama}, \citenamefont {Yoshimura},\ and\ \citenamefont {Ito}}]{naga25}%
  \BibitemOpen
  \bibfield  {author} {\bibinfo {author} {\bibfnamefont {R.}~\bibnamefont
  {Nagayama}}, \bibinfo {author} {\bibfnamefont {K.}~\bibnamefont
  {Yoshimura}},\ and\ \bibinfo {author} {\bibfnamefont {S.}~\bibnamefont
  {Ito}},\ }\bibfield  {title} {\bibinfo {title} {Infinite variety of
  thermodynamic speed limits with general activities},\ }\href
  {https://doi.org/10.1103/PhysRevResearch.7.013307} {\bibfield  {journal}
  {\bibinfo  {journal} {Physical Review Research}\ }\textbf {\bibinfo {volume}
  {7}},\ \bibinfo {pages} {013307} (\bibinfo {year} {2025})}\BibitemShut
  {NoStop}%
\bibitem [{\citenamefont {Dechant}\ and\ \citenamefont {Sasa}(2020)}]{dech20}%
  \BibitemOpen
  \bibfield  {author} {\bibinfo {author} {\bibfnamefont {A.}~\bibnamefont
  {Dechant}}\ and\ \bibinfo {author} {\bibfnamefont {S.-i.}\ \bibnamefont
  {Sasa}},\ }\bibfield  {title} {\bibinfo {title} {Fluctuation--response
  inequality out of equilibrium},\ }\href
  {https://doi.org/10.1073/pnas.1918386117} {\bibfield  {journal} {\bibinfo
  {journal} {Proc.\ Natl.\ Acad.\ Sci.\ U.S.A.}\ }\textbf {\bibinfo {volume}
  {117}},\ \bibinfo {pages} {6430} (\bibinfo {year} {2020})}\BibitemShut
  {NoStop}%
\bibitem [{\citenamefont {Owen}\ \emph {et~al.}(2020)\citenamefont {Owen},
  \citenamefont {Gingrich},\ and\ \citenamefont {Horowitz}}]{owen20}%
  \BibitemOpen
  \bibfield  {author} {\bibinfo {author} {\bibfnamefont {J.~A.}\ \bibnamefont
  {Owen}}, \bibinfo {author} {\bibfnamefont {T.~R.}\ \bibnamefont {Gingrich}},\
  and\ \bibinfo {author} {\bibfnamefont {J.~M.}\ \bibnamefont {Horowitz}},\
  }\bibfield  {title} {\bibinfo {title} {Universal {{Thermodynamic Bounds}} on
  {{Nonequilibrium Response}} with {{Biochemical Applications}}},\ }\href
  {https://doi.org/10.1103/PhysRevX.10.011066} {\bibfield  {journal} {\bibinfo
  {journal} {Physical Review X}\ }\textbf {\bibinfo {volume} {10}},\ \bibinfo
  {pages} {011066} (\bibinfo {year} {2020})}\BibitemShut {NoStop}%
\bibitem [{\citenamefont {Aslyamov}\ and\ \citenamefont
  {Esposito}(2024)}]{asly24}%
  \BibitemOpen
  \bibfield  {author} {\bibinfo {author} {\bibfnamefont {T.}~\bibnamefont
  {Aslyamov}}\ and\ \bibinfo {author} {\bibfnamefont {M.}~\bibnamefont
  {Esposito}},\ }\bibfield  {title} {\bibinfo {title} {Nonequilibrium
  {{Response}} for {{Markov Jump Processes}}: {{Exact Results}} and {{Tight
  Bounds}}},\ }\href {https://doi.org/10.1103/PhysRevLett.132.037101}
  {\bibfield  {journal} {\bibinfo  {journal} {Phys.\ Rev.\ Lett.}\ }\textbf
  {\bibinfo {volume} {132}},\ \bibinfo {pages} {037101} (\bibinfo {year}
  {2024})}\BibitemShut {NoStop}%
\bibitem [{\citenamefont {Ohga}\ \emph {et~al.}(2023)\citenamefont {Ohga},
  \citenamefont {Ito},\ and\ \citenamefont {Kolchinsky}}]{ohga23}%
  \BibitemOpen
  \bibfield  {author} {\bibinfo {author} {\bibfnamefont {N.}~\bibnamefont
  {Ohga}}, \bibinfo {author} {\bibfnamefont {S.}~\bibnamefont {Ito}},\ and\
  \bibinfo {author} {\bibfnamefont {A.}~\bibnamefont {Kolchinsky}},\ }\bibfield
   {title} {\bibinfo {title} {Thermodynamic bound on the asymmetry of
  cross-correlations},\ }\href {https://doi.org/10.1103/PhysRevLett.131.077101}
  {\bibfield  {journal} {\bibinfo  {journal} {Phys. Rev. Lett.}\ }\textbf
  {\bibinfo {volume} {131}},\ \bibinfo {pages} {077101} (\bibinfo {year}
  {2023})}\BibitemShut {NoStop}%
\bibitem [{\citenamefont {Shiraishi}(2023{\natexlab{b}})}]{shir23b}%
  \BibitemOpen
  \bibfield  {author} {\bibinfo {author} {\bibfnamefont {N.}~\bibnamefont
  {Shiraishi}},\ }\bibfield  {title} {\bibinfo {title} {Entropy production
  limits all fluctuation oscillations},\ }\href
  {https://doi.org/10.1103/PhysRevE.108.L042103} {\bibfield  {journal}
  {\bibinfo  {journal} {Phys. Rev. E}\ }\textbf {\bibinfo {volume} {108}},\
  \bibinfo {pages} {L042103} (\bibinfo {year}
  {2023}{\natexlab{b}})}\BibitemShut {NoStop}%
\bibitem [{\citenamefont {Skinner}\ and\ \citenamefont
  {Dunkel}(2021)}]{skin21a}%
  \BibitemOpen
  \bibfield  {author} {\bibinfo {author} {\bibfnamefont {D.~J.}\ \bibnamefont
  {Skinner}}\ and\ \bibinfo {author} {\bibfnamefont {J.}~\bibnamefont
  {Dunkel}},\ }\bibfield  {title} {\bibinfo {title} {Estimating entropy
  production from waiting time distributions},\ }\href
  {https://link.aps.org/doi/10.1103/PhysRevLett.127.198101} {\bibfield
  {journal} {\bibinfo  {journal} {Phys.\ Rev.\ Lett.}\ }\textbf {\bibinfo
  {volume} {127}},\ \bibinfo {pages} {198101} (\bibinfo {year}
  {2021})}\BibitemShut {NoStop}%
\bibitem [{\citenamefont {Dechant}\ \emph {et~al.}(2023)\citenamefont
  {Dechant}, \citenamefont {Garnier-Brun},\ and\ \citenamefont
  {Sasa}}]{dech23}%
  \BibitemOpen
  \bibfield  {author} {\bibinfo {author} {\bibfnamefont {A.}~\bibnamefont
  {Dechant}}, \bibinfo {author} {\bibfnamefont {J.}~\bibnamefont
  {Garnier-Brun}},\ and\ \bibinfo {author} {\bibfnamefont {S.-i.}\ \bibnamefont
  {Sasa}},\ }\bibfield  {title} {\bibinfo {title} {Thermodynamic bounds on
  correlation times},\ }\href {https://doi.org/10.1103/PhysRevLett.131.167101}
  {\bibfield  {journal} {\bibinfo  {journal} {Phys. Rev. Lett.}\ }\textbf
  {\bibinfo {volume} {131}},\ \bibinfo {pages} {167101} (\bibinfo {year}
  {2023})}\BibitemShut {NoStop}%
\bibitem [{\citenamefont {Pietzonka}\ and\ \citenamefont
  {Coghi}(2024)}]{piet24}%
  \BibitemOpen
  \bibfield  {author} {\bibinfo {author} {\bibfnamefont {P.}~\bibnamefont
  {Pietzonka}}\ and\ \bibinfo {author} {\bibfnamefont {F.}~\bibnamefont
  {Coghi}},\ }\bibfield  {title} {\bibinfo {title} {Thermodynamic cost for
  precision of general counting observables},\ }\href
  {https://doi.org/10.1103/PhysRevE.109.064128} {\bibfield  {journal} {\bibinfo
   {journal} {Phys. Rev. E}\ }\textbf {\bibinfo {volume} {109}},\ \bibinfo
  {pages} {064128} (\bibinfo {year} {2024})}\BibitemShut {NoStop}%
\bibitem [{\citenamefont {Esposito}(2012)}]{espo12}%
  \BibitemOpen
  \bibfield  {author} {\bibinfo {author} {\bibfnamefont {M.}~\bibnamefont
  {Esposito}},\ }\bibfield  {title} {\bibinfo {title} {Stochastic
  thermodynamics under coarse graining},\ }\href
  {https://doi.org/10.1103/PhysRevE.85.041125} {\bibfield  {journal} {\bibinfo
  {journal} {Physical Review E}\ }\textbf {\bibinfo {volume} {85}},\ \bibinfo
  {pages} {041125} (\bibinfo {year} {2012})}\BibitemShut {NoStop}%
\bibitem [{\citenamefont {Bo}\ and\ \citenamefont {Celani}(2017)}]{bo17}%
  \BibitemOpen
  \bibfield  {author} {\bibinfo {author} {\bibfnamefont {S.}~\bibnamefont
  {Bo}}\ and\ \bibinfo {author} {\bibfnamefont {A.}~\bibnamefont {Celani}},\
  }\bibfield  {title} {\bibinfo {title} {Multiple-scale stochastic processes:
  {{Decimation}}, averaging and beyond},\ }\href
  {https://doi.org/10.1016/j.physrep.2016.12.003} {\bibfield  {journal}
  {\bibinfo  {journal} {Physics Reports}\ }\textbf {\bibinfo {volume} {670}},\
  \bibinfo {pages} {1} (\bibinfo {year} {2017})}\BibitemShut {NoStop}%
\bibitem [{\citenamefont {Shiraishi}\ and\ \citenamefont
  {Sagawa}(2015)}]{shir15}%
  \BibitemOpen
  \bibfield  {author} {\bibinfo {author} {\bibfnamefont {N.}~\bibnamefont
  {Shiraishi}}\ and\ \bibinfo {author} {\bibfnamefont {T.}~\bibnamefont
  {Sagawa}},\ }\bibfield  {title} {\bibinfo {title} {Fluctuation theorem for
  partially masked nonequilibrium dynamics},\ }\href
  {https://doi.org/10.1103/PhysRevE.91.012130} {\bibfield  {journal} {\bibinfo
  {journal} {Physical Review E}\ }\textbf {\bibinfo {volume} {91}},\ \bibinfo
  {pages} {012130} (\bibinfo {year} {2015})}\BibitemShut {NoStop}%
\bibitem [{\citenamefont {Polettini}\ and\ \citenamefont
  {Esposito}(2019)}]{pole19}%
  \BibitemOpen
  \bibfield  {author} {\bibinfo {author} {\bibfnamefont {M.}~\bibnamefont
  {Polettini}}\ and\ \bibinfo {author} {\bibfnamefont {M.}~\bibnamefont
  {Esposito}},\ }\bibfield  {title} {\bibinfo {title} {Effective
  {{Fluctuation}} and {{Response Theory}}},\ }\href
  {https://doi.org/10.1007/s10955-019-02291-7} {\bibfield  {journal} {\bibinfo
  {journal} {Journal of Statistical Physics}\ }\textbf {\bibinfo {volume}
  {176}},\ \bibinfo {pages} {94} (\bibinfo {year} {2019})}\BibitemShut
  {NoStop}%
\bibitem [{\citenamefont {Mart{\'i}nez}\ \emph {et~al.}(2019)\citenamefont
  {Mart{\'i}nez}, \citenamefont {Bisker}, \citenamefont {Horowitz},\ and\
  \citenamefont {Parrondo}}]{mart19}%
  \BibitemOpen
  \bibfield  {author} {\bibinfo {author} {\bibfnamefont {I.~A.}\ \bibnamefont
  {Mart{\'i}nez}}, \bibinfo {author} {\bibfnamefont {G.}~\bibnamefont
  {Bisker}}, \bibinfo {author} {\bibfnamefont {J.~M.}\ \bibnamefont
  {Horowitz}},\ and\ \bibinfo {author} {\bibfnamefont {J.~M.~R.}\ \bibnamefont
  {Parrondo}},\ }\bibfield  {title} {\bibinfo {title} {Inferring broken
  detailed balance in the absence of observable currents},\ }\href
  {https://doi.org/10.1038/s41467-019-11051-w} {\bibfield  {journal} {\bibinfo
  {journal} {Nature Communications}\ }\textbf {\bibinfo {volume} {10}},\
  \bibinfo {pages} {3542} (\bibinfo {year} {2019})}\BibitemShut {NoStop}%
\bibitem [{\citenamefont {{van der Meer}}\ \emph {et~al.}(2022)\citenamefont
  {{van der Meer}}, \citenamefont {Ertel},\ and\ \citenamefont
  {Seifert}}]{vdm22}%
  \BibitemOpen
  \bibfield  {author} {\bibinfo {author} {\bibfnamefont {J.}~\bibnamefont {{van
  der Meer}}}, \bibinfo {author} {\bibfnamefont {B.}~\bibnamefont {Ertel}},\
  and\ \bibinfo {author} {\bibfnamefont {U.}~\bibnamefont {Seifert}},\
  }\bibfield  {title} {\bibinfo {title} {Thermodynamic inference in partially
  accessible markov networks: A unifying perspective from transition-based
  waiting time distributions},\ }\href
  {https://doi.org/https://doi.org/10.1103/PhysRevX.12.031025} {\bibfield
  {journal} {\bibinfo  {journal} {Phys. Rev. X}\ }\textbf {\bibinfo {volume}
  {12}},\ \bibinfo {pages} {031025} (\bibinfo {year} {2022})}\BibitemShut
  {NoStop}%
\bibitem [{\citenamefont {Harunari}\ \emph {et~al.}(2022)\citenamefont
  {Harunari}, \citenamefont {Dutta}, \citenamefont {Polettini},\ and\
  \citenamefont {Roldan}}]{haru22}%
  \BibitemOpen
  \bibfield  {author} {\bibinfo {author} {\bibfnamefont {P.}~\bibnamefont
  {Harunari}}, \bibinfo {author} {\bibfnamefont {A.}~\bibnamefont {Dutta}},
  \bibinfo {author} {\bibfnamefont {M.}~\bibnamefont {Polettini}},\ and\
  \bibinfo {author} {\bibfnamefont {E.}~\bibnamefont {Roldan}},\ }\bibfield
  {title} {\bibinfo {title} {What to learn from a few visible transitions’
  statistics?},\ }\href
  {https://doi.org/https://doi.org/10.1103/PhysRevX.12.041026} {\bibfield
  {journal} {\bibinfo  {journal} {Phys. Rev. X}\ }\textbf {\bibinfo {volume}
  {12}},\ \bibinfo {pages} {041026} (\bibinfo {year} {2022})}\BibitemShut
  {NoStop}%
\bibitem [{\citenamefont {Gomez-Marin}\ \emph {et~al.}(2008)\citenamefont
  {Gomez-Marin}, \citenamefont {Parrondo},\ and\ \citenamefont {{Van den
  Broeck}}}]{gome08a}%
  \BibitemOpen
  \bibfield  {author} {\bibinfo {author} {\bibfnamefont {A.}~\bibnamefont
  {Gomez-Marin}}, \bibinfo {author} {\bibfnamefont {J.~M.~R.}\ \bibnamefont
  {Parrondo}},\ and\ \bibinfo {author} {\bibfnamefont {C.}~\bibnamefont {{Van
  den Broeck}}},\ }\bibfield  {title} {\bibinfo {title} {Lower bounds on
  dissipation upon coarse-graining},\ }\href
  {https://doi.org/10.1103/PhysRevE.78.011107} {\bibfield  {journal} {\bibinfo
  {journal} {Phys. Rev. E}\ }\textbf {\bibinfo {volume} {78}},\ \bibinfo
  {pages} {011107} (\bibinfo {year} {2008})}\BibitemShut {NoStop}%
\bibitem [{\citenamefont {Roldan}\ and\ \citenamefont
  {Parrondo}(2010)}]{rold10}%
  \BibitemOpen
  \bibfield  {author} {\bibinfo {author} {\bibfnamefont {E.}~\bibnamefont
  {Roldan}}\ and\ \bibinfo {author} {\bibfnamefont {J.~M.~R.}\ \bibnamefont
  {Parrondo}},\ }\bibfield  {title} {\bibinfo {title} {Estimating dissipation
  from single stationary trajectories},\ }\href
  {https://doi.org/10.1103/PhysRevLett.105.150607} {\bibfield  {journal}
  {\bibinfo  {journal} {Phys. Rev. Lett.}\ }\textbf {\bibinfo {volume} {105}},\
  \bibinfo {pages} {150607} (\bibinfo {year} {2010})}\BibitemShut {NoStop}%
\bibitem [{\citenamefont {Muy}\ \emph {et~al.}(2013)\citenamefont {Muy},
  \citenamefont {Kundu},\ and\ \citenamefont {Lacoste}}]{muy13}%
  \BibitemOpen
  \bibfield  {author} {\bibinfo {author} {\bibfnamefont {S.}~\bibnamefont
  {Muy}}, \bibinfo {author} {\bibfnamefont {A.}~\bibnamefont {Kundu}},\ and\
  \bibinfo {author} {\bibfnamefont {D.}~\bibnamefont {Lacoste}},\ }\bibfield
  {title} {\bibinfo {title} {Non-invasive estimation of dissipation from
  non-equilibrium fluctuations in chemical reactions},\ }\href
  {https://doi.org/10.1063/1.4821760} {\bibfield  {journal} {\bibinfo
  {journal} {J.\ Chem.\ Phys.}\ }\textbf {\bibinfo {volume} {139}},\ \bibinfo
  {pages} {124109} (\bibinfo {year} {2013})}\BibitemShut {NoStop}%
\bibitem [{\citenamefont {Harunari}\ \emph {et~al.}(2023)\citenamefont
  {Harunari}, \citenamefont {Garilli},\ and\ \citenamefont
  {Polettini}}]{haru23a}%
  \BibitemOpen
  \bibfield  {author} {\bibinfo {author} {\bibfnamefont {P.~E.}\ \bibnamefont
  {Harunari}}, \bibinfo {author} {\bibfnamefont {A.}~\bibnamefont {Garilli}},\
  and\ \bibinfo {author} {\bibfnamefont {M.}~\bibnamefont {Polettini}},\
  }\bibfield  {title} {\bibinfo {title} {Beat of a current},\ }\href
  {https://doi.org/10.1103/PhysRevE.107.L042105} {\bibfield  {journal}
  {\bibinfo  {journal} {Physical Review E}\ }\textbf {\bibinfo {volume}
  {107}},\ \bibinfo {pages} {L042105} (\bibinfo {year} {2023})}\BibitemShut
  {NoStop}%
\bibitem [{\citenamefont {Deg{\"u}nther}\ \emph
  {et~al.}(2024{\natexlab{a}})\citenamefont {Deg{\"u}nther}, \citenamefont {Van
  Der~Meer},\ and\ \citenamefont {Seifert}}]{degu24}%
  \BibitemOpen
  \bibfield  {author} {\bibinfo {author} {\bibfnamefont {J.}~\bibnamefont
  {Deg{\"u}nther}}, \bibinfo {author} {\bibfnamefont {J.}~\bibnamefont {Van
  Der~Meer}},\ and\ \bibinfo {author} {\bibfnamefont {U.}~\bibnamefont
  {Seifert}},\ }\bibfield  {title} {\bibinfo {title} {Fluctuating entropy
  production on the coarse-grained level: {{Inference}} and localization of
  irreversibility},\ }\href {https://doi.org/10.1103/PhysRevResearch.6.023175}
  {\bibfield  {journal} {\bibinfo  {journal} {Physical Review Research}\
  }\textbf {\bibinfo {volume} {6}},\ \bibinfo {pages} {023175} (\bibinfo {year}
  {2024}{\natexlab{a}})}\BibitemShut {NoStop}%
\bibitem [{\citenamefont {Deg{\"u}nther}\ \emph
  {et~al.}(2024{\natexlab{b}})\citenamefont {Deg{\"u}nther}, \citenamefont
  {{van der Meer}},\ and\ \citenamefont {Seifert}}]{degu24b}%
  \BibitemOpen
  \bibfield  {author} {\bibinfo {author} {\bibfnamefont {J.}~\bibnamefont
  {Deg{\"u}nther}}, \bibinfo {author} {\bibfnamefont {J.}~\bibnamefont {{van
  der Meer}}},\ and\ \bibinfo {author} {\bibfnamefont {U.}~\bibnamefont
  {Seifert}},\ }\bibfield  {title} {\bibinfo {title} {General theory for
  localizing the where and when of entropy production meets single-molecule
  experiments},\ }\href {https://doi.org/10.1073/pnas.2405371121} {\bibfield
  {journal} {\bibinfo  {journal} {Proc.\ Natl.\ Acad.\ Sci.\ U.S.A.}\ }\textbf
  {\bibinfo {volume} {121}},\ \bibinfo {pages} {e2405371121} (\bibinfo {year}
  {2024}{\natexlab{b}})}\BibitemShut {NoStop}%
\bibitem [{\citenamefont {Hartich}\ and\ \citenamefont {Godec}(2021)}]{hart21}%
  \BibitemOpen
  \bibfield  {author} {\bibinfo {author} {\bibfnamefont {D.}~\bibnamefont
  {Hartich}}\ and\ \bibinfo {author} {\bibfnamefont {A.}~\bibnamefont
  {Godec}},\ }\bibfield  {title} {\bibinfo {title} {Emergent {{Memory}} and
  {{Kinetic Hysteresis}} in {{Strongly Driven Networks}}},\ }\href
  {https://doi.org/10.1103/PhysRevX.11.041047} {\bibfield  {journal} {\bibinfo
  {journal} {Physical Review X}\ }\textbf {\bibinfo {volume} {11}},\ \bibinfo
  {pages} {041047} (\bibinfo {year} {2021})}\BibitemShut {NoStop}%
\bibitem [{\citenamefont {Blom}\ \emph {et~al.}(2024)\citenamefont {Blom},
  \citenamefont {Song}, \citenamefont {Vouga}, \citenamefont {Godec},\ and\
  \citenamefont {Makarov}}]{blom24a}%
  \BibitemOpen
  \bibfield  {author} {\bibinfo {author} {\bibfnamefont {K.}~\bibnamefont
  {Blom}}, \bibinfo {author} {\bibfnamefont {K.}~\bibnamefont {Song}}, \bibinfo
  {author} {\bibfnamefont {E.}~\bibnamefont {Vouga}}, \bibinfo {author}
  {\bibfnamefont {A.}~\bibnamefont {Godec}},\ and\ \bibinfo {author}
  {\bibfnamefont {D.~E.}\ \bibnamefont {Makarov}},\ }\bibfield  {title}
  {\bibinfo {title} {Milestoning estimators of dissipation in systems observed
  at a coarse resolution},\ }\href {https://doi.org/10.1073/pnas.2318333121}
  {\bibfield  {journal} {\bibinfo  {journal} {Proc.\ Natl.\ Acad.\ Sci.\
  U.S.A.}\ }\textbf {\bibinfo {volume} {121}},\ \bibinfo {pages} {e2318333121}
  (\bibinfo {year} {2024})}\BibitemShut {NoStop}%
\bibitem [{\citenamefont {{van der Meer}}\ \emph {et~al.}(2023)\citenamefont
  {{van der Meer}}, \citenamefont {Deg\"unther},\ and\ \citenamefont
  {Seifert}}]{vdm23}%
  \BibitemOpen
  \bibfield  {author} {\bibinfo {author} {\bibfnamefont {J.}~\bibnamefont {{van
  der Meer}}}, \bibinfo {author} {\bibfnamefont {J.}~\bibnamefont
  {Deg\"unther}},\ and\ \bibinfo {author} {\bibfnamefont {U.}~\bibnamefont
  {Seifert}},\ }\bibfield  {title} {\bibinfo {title} {Time-resolved statistics
  of snippets as general framework for model-free entropy estimators},\ }\href
  {https://doi.org/10.1103/PhysRevLett.130.257101} {\bibfield  {journal}
  {\bibinfo  {journal} {Phys. Rev. Lett.}\ }\textbf {\bibinfo {volume} {130}},\
  \bibinfo {pages} {257101} (\bibinfo {year} {2023})}\BibitemShut {NoStop}%
\bibitem [{\citenamefont {Igoshin}\ \emph
  {et~al.}(2025{\natexlab{a}})\citenamefont {Igoshin}, \citenamefont
  {Kolomeisky},\ and\ \citenamefont {Makarov}}]{igos25a}%
  \BibitemOpen
  \bibfield  {author} {\bibinfo {author} {\bibfnamefont {O.~A.}\ \bibnamefont
  {Igoshin}}, \bibinfo {author} {\bibfnamefont {A.~B.}\ \bibnamefont
  {Kolomeisky}},\ and\ \bibinfo {author} {\bibfnamefont {D.~E.}\ \bibnamefont
  {Makarov}},\ }\bibfield  {title} {\bibinfo {title} {Coarse-{{Graining
  Chemical Networks}} by {{Trimming}} to {{Preserve Energy Dissipation}}},\
  }\href {https://doi.org/10.1021/acs.jpclett.4c03372} {\bibfield  {journal}
  {\bibinfo  {journal} {The Journal of Physical Chemistry Letters}\ }\textbf
  {\bibinfo {volume} {16}},\ \bibinfo {pages} {1229} (\bibinfo {year}
  {2025}{\natexlab{a}})}\BibitemShut {NoStop}%
\bibitem [{\citenamefont {Li}\ and\ \citenamefont {Kolomeisky}(2013)}]{li13}%
  \BibitemOpen
  \bibfield  {author} {\bibinfo {author} {\bibfnamefont {X.}~\bibnamefont
  {Li}}\ and\ \bibinfo {author} {\bibfnamefont {A.~B.}\ \bibnamefont
  {Kolomeisky}},\ }\bibfield  {title} {\bibinfo {title} {Mechanisms and
  topology determination of complex chemical and biological network systems
  from first-passage theoretical approach},\ }\href
  {https://doi.org/10.1063/1.4824392} {\bibfield  {journal} {\bibinfo
  {journal} {J.\ Chem.\ Phys.}\ }\textbf {\bibinfo {volume} {139}},\ \bibinfo
  {pages} {144106} (\bibinfo {year} {2013})}\BibitemShut {NoStop}%
\bibitem [{\citenamefont {Berezhkovskii}\ \emph {et~al.}(2006)\citenamefont
  {Berezhkovskii}, \citenamefont {Hummer},\ and\ \citenamefont
  {Bezrukov}}]{bere06}%
  \BibitemOpen
  \bibfield  {author} {\bibinfo {author} {\bibfnamefont {A.~M.}\ \bibnamefont
  {Berezhkovskii}}, \bibinfo {author} {\bibfnamefont {G.}~\bibnamefont
  {Hummer}},\ and\ \bibinfo {author} {\bibfnamefont {S.~M.}\ \bibnamefont
  {Bezrukov}},\ }\bibfield  {title} {\bibinfo {title} {Identity of
  distributions of direct uphill and downhill translocation times for particles
  traversing membrane channels},\ }\href
  {https://doi.org/10.1103/PhysRevLett.97.020601} {\bibfield  {journal}
  {\bibinfo  {journal} {Phys. Rev. Lett.}\ }\textbf {\bibinfo {volume} {97}},\
  \bibinfo {pages} {020601} (\bibinfo {year} {2006})}\BibitemShut {NoStop}%
\bibitem [{\citenamefont {Berezhkovskii}\ and\ \citenamefont
  {Makarov}(2019)}]{bere19}%
  \BibitemOpen
  \bibfield  {author} {\bibinfo {author} {\bibfnamefont {A.~M.}\ \bibnamefont
  {Berezhkovskii}}\ and\ \bibinfo {author} {\bibfnamefont {D.~E.}\ \bibnamefont
  {Makarov}},\ }\bibfield  {title} {\bibinfo {title} {On the forward/backward
  symmetry of transition path time distributions in nonequilibrium systems},\
  }\href {https://doi.org/10.1063/1.5109293} {\bibfield  {journal} {\bibinfo
  {journal} {J.\ Chem.\ Phys.}\ }\textbf {\bibinfo {volume} {151}},\ \bibinfo
  {pages} {065102} (\bibinfo {year} {2019})}\BibitemShut {NoStop}%
\bibitem [{\citenamefont {Gladrow}\ \emph {et~al.}(2019)\citenamefont
  {Gladrow}, \citenamefont {{Ribezzi-Crivellari}}, \citenamefont {Ritort},\
  and\ \citenamefont {Keyser}}]{glad19}%
  \BibitemOpen
  \bibfield  {author} {\bibinfo {author} {\bibfnamefont {J.}~\bibnamefont
  {Gladrow}}, \bibinfo {author} {\bibfnamefont {M.}~\bibnamefont
  {{Ribezzi-Crivellari}}}, \bibinfo {author} {\bibfnamefont {F.}~\bibnamefont
  {Ritort}},\ and\ \bibinfo {author} {\bibfnamefont {U.~F.}\ \bibnamefont
  {Keyser}},\ }\bibfield  {title} {\bibinfo {title} {Experimental evidence of
  symmetry breaking of transition-path times},\ }\href
  {https://doi.org/10.1038/s41467-018-07873-9} {\bibfield  {journal} {\bibinfo
  {journal} {Nature Communications}\ }\textbf {\bibinfo {volume} {10}},\
  \bibinfo {pages} {55} (\bibinfo {year} {2019})}\BibitemShut {NoStop}%
\bibitem [{\citenamefont {Bowman}\ \emph {et~al.}(2014)\citenamefont {Bowman},
  \citenamefont {Pande},\ and\ \citenamefont {Noé}}]{bowm14}%
  \BibitemOpen
  \bibinfo {editor} {\bibfnamefont {G.~R.}\ \bibnamefont {Bowman}}, \bibinfo
  {editor} {\bibfnamefont {V.~S.}\ \bibnamefont {Pande}},\ and\ \bibinfo
  {editor} {\bibfnamefont {F.}~\bibnamefont {Noé}},\ eds.,\ \href
  {https://doi.org/10.1007/978-94-007-7606-7} {\emph {\bibinfo {title} {An
  Introduction to Markov State Models and Their Application to Long Timescale
  Molecular Simulation}}}\ (\bibinfo  {publisher} {Springer Dordrecht},\
  \bibinfo {year} {2014})\BibitemShut {NoStop}%
\bibitem [{\citenamefont {Voll{mar}}\ \emph {et~al.}(2024)\citenamefont
  {Voll{mar}}, \citenamefont {Schimpf}, \citenamefont {Hermann},\ and\
  \citenamefont {Hugel}}]{voll24a}%
  \BibitemOpen
  \bibfield  {author} {\bibinfo {author} {\bibfnamefont {L.}~\bibnamefont
  {Voll{mar}}}, \bibinfo {author} {\bibfnamefont {J.}~\bibnamefont {Schimpf}},
  \bibinfo {author} {\bibfnamefont {B.}~\bibnamefont {Hermann}},\ and\ \bibinfo
  {author} {\bibfnamefont {T.}~\bibnamefont {Hugel}},\ }\bibfield  {title}
  {\bibinfo {title} {Cochaperones convey the energy of {{ATP}} hydrolysis for
  directional action of {{Hsp90}}},\ }\href
  {https://doi.org/10.1038/s41467-024-44847-6} {\bibfield  {journal} {\bibinfo
  {journal} {Nature Communications}\ }\textbf {\bibinfo {volume} {15}},\
  \bibinfo {pages} {569} (\bibinfo {year} {2024})}\BibitemShut {NoStop}%
\bibitem [{\citenamefont {Vollmar}\ \emph {et~al.}(2024)\citenamefont
  {Vollmar}, \citenamefont {Bebon}, \citenamefont {Schimpf}, \citenamefont
  {Flietel}, \citenamefont {Celiksoy}, \citenamefont {Sönnichsen},
  \citenamefont {Godec},\ and\ \citenamefont {Hugel}}]{voll24}%
  \BibitemOpen
  \bibfield  {author} {\bibinfo {author} {\bibfnamefont {L.}~\bibnamefont
  {Vollmar}}, \bibinfo {author} {\bibfnamefont {R.}~\bibnamefont {Bebon}},
  \bibinfo {author} {\bibfnamefont {J.}~\bibnamefont {Schimpf}}, \bibinfo
  {author} {\bibfnamefont {B.}~\bibnamefont {Flietel}}, \bibinfo {author}
  {\bibfnamefont {S.}~\bibnamefont {Celiksoy}}, \bibinfo {author}
  {\bibfnamefont {C.}~\bibnamefont {Sönnichsen}}, \bibinfo {author}
  {\bibfnamefont {A.}~\bibnamefont {Godec}},\ and\ \bibinfo {author}
  {\bibfnamefont {T.}~\bibnamefont {Hugel}},\ }\bibfield  {title} {\bibinfo
  {title} {Model-free inference of memory in conformational dynamics of a
  multi-domain protein},\ }\href {https://doi.org/10.1088/1751-8121/ad6d1e}
  {\bibfield  {journal} {\bibinfo  {journal} {Journal of Physics A:
  Mathematical and Theoretical}\ }\textbf {\bibinfo {volume} {57}},\ \bibinfo
  {pages} {365001} (\bibinfo {year} {2024})}\BibitemShut {NoStop}%
\bibitem [{\citenamefont {Neri}\ and\ \citenamefont
  {Polettini}(2023)}]{neri23}%
  \BibitemOpen
  \bibfield  {author} {\bibinfo {author} {\bibfnamefont {I.}~\bibnamefont
  {Neri}}\ and\ \bibinfo {author} {\bibfnamefont {M.}~\bibnamefont
  {Polettini}},\ }\bibfield  {title} {\bibinfo {title} {{Extreme value
  statistics of edge currents in Markov jump processes and their use for
  entropy production estimation}},\ }\href
  {https://doi.org/10.21468/SciPostPhys.14.5.131} {\bibfield  {journal}
  {\bibinfo  {journal} {SciPost Phys.}\ }\textbf {\bibinfo {volume} {14}},\
  \bibinfo {pages} {131} (\bibinfo {year} {2023})}\BibitemShut {NoStop}%
\bibitem [{\citenamefont {J\"ulicher}\ \emph {et~al.}(1997)\citenamefont
  {J\"ulicher}, \citenamefont {Ajdari},\ and\ \citenamefont {Prost}}]{juel97}%
  \BibitemOpen
  \bibfield  {author} {\bibinfo {author} {\bibfnamefont {F.}~\bibnamefont
  {J\"ulicher}}, \bibinfo {author} {\bibfnamefont {A.}~\bibnamefont {Ajdari}},\
  and\ \bibinfo {author} {\bibfnamefont {J.}~\bibnamefont {Prost}},\ }\bibfield
   {title} {\bibinfo {title} {Modeling molecular motors},\ }\href
  {https://doi.org/10.1103/RevModPhys.69.1269} {\bibfield  {journal} {\bibinfo
  {journal} {Rev. Mod. Phys.}\ }\textbf {\bibinfo {volume} {69}},\ \bibinfo
  {pages} {1269} (\bibinfo {year} {1997})}\BibitemShut {NoStop}%
\bibitem [{\citenamefont {Gaspard}\ and\ \citenamefont
  {Gerritsma}(2007)}]{gasp07}%
  \BibitemOpen
  \bibfield  {author} {\bibinfo {author} {\bibfnamefont {P.}~\bibnamefont
  {Gaspard}}\ and\ \bibinfo {author} {\bibfnamefont {E.}~\bibnamefont
  {Gerritsma}},\ }\bibfield  {title} {\bibinfo {title} {The stochastic
  chemomechanics of the {F}$_1$-{ATP}ase molecular motor},\ }\href
  {https://doi.org/10.1016/j.jtbi.2007.03.034} {\bibfield  {journal} {\bibinfo
  {journal} {J.\ Theor.\ Biol.}\ }\textbf {\bibinfo {volume} {247}},\ \bibinfo
  {pages} {672} (\bibinfo {year} {2007})}\BibitemShut {NoStop}%
\bibitem [{\citenamefont {Zimmermann}\ and\ \citenamefont
  {Seifert}(2012)}]{zimm12}%
  \BibitemOpen
  \bibfield  {author} {\bibinfo {author} {\bibfnamefont {E.}~\bibnamefont
  {Zimmermann}}\ and\ \bibinfo {author} {\bibfnamefont {U.}~\bibnamefont
  {Seifert}},\ }\bibfield  {title} {\bibinfo {title} {Efficiency of a molecular
  motor: A generic hybrid model applied to the {F}$_1$-{ATP}ase},\ }\href
  {https://doi.org/10.1088/1367-2630/14/10/103023} {\bibfield  {journal}
  {\bibinfo  {journal} {New\ J.\ Phys.}\ }\textbf {\bibinfo {volume} {14}},\
  \bibinfo {pages} {103023} (\bibinfo {year} {2012})}\BibitemShut {NoStop}%
\bibitem [{\citenamefont {Maier}\ \emph {et~al.}(2024)\citenamefont {Maier},
  \citenamefont {Deg{\"u}nther}, \citenamefont {{van der Meer}},\ and\
  \citenamefont {Seifert}}]{maie24}%
  \BibitemOpen
  \bibfield  {author} {\bibinfo {author} {\bibfnamefont {A.~M.}\ \bibnamefont
  {Maier}}, \bibinfo {author} {\bibfnamefont {J.}~\bibnamefont
  {Deg{\"u}nther}}, \bibinfo {author} {\bibfnamefont {J.}~\bibnamefont {{van
  der Meer}}},\ and\ \bibinfo {author} {\bibfnamefont {U.}~\bibnamefont
  {Seifert}},\ }\bibfield  {title} {\bibinfo {title} {Inferring kinetics and
  entropy production from observable transitions in partially accessible,
  periodically driven markov networks},\ }\href
  {https://doi.org/10.1007/s10955-024-03315-7} {\bibfield  {journal} {\bibinfo
  {journal} {Journal of Statistical Physics}\ }\textbf {\bibinfo {volume}
  {191}},\ \bibinfo {pages} {104} (\bibinfo {year} {2024})}\BibitemShut
  {NoStop}%
\bibitem [{\citenamefont {Harunari}\ \emph {et~al.}(2024)\citenamefont
  {Harunari}, \citenamefont {Fiore},\ and\ \citenamefont {Barato}}]{haru24b}%
  \BibitemOpen
  \bibfield  {author} {\bibinfo {author} {\bibfnamefont {P.~E.}\ \bibnamefont
  {Harunari}}, \bibinfo {author} {\bibfnamefont {C.~E.}\ \bibnamefont
  {Fiore}},\ and\ \bibinfo {author} {\bibfnamefont {A.~C.}\ \bibnamefont
  {Barato}},\ }\bibfield  {title} {\bibinfo {title} {Inference of entropy
  production for periodically driven systems},\ }\href
  {https://doi.org/10.1103/PhysRevE.110.064126} {\bibfield  {journal} {\bibinfo
   {journal} {Physical Review E}\ }\textbf {\bibinfo {volume} {110}},\ \bibinfo
  {pages} {064126} (\bibinfo {year} {2024})}\BibitemShut {NoStop}%
\bibitem [{\citenamefont {Godec}\ and\ \citenamefont {Makarov}(2023)}]{gode23}%
  \BibitemOpen
  \bibfield  {author} {\bibinfo {author} {\bibfnamefont {A.}~\bibnamefont
  {Godec}}\ and\ \bibinfo {author} {\bibfnamefont {D.~E.}\ \bibnamefont
  {Makarov}},\ }\bibfield  {title} {\bibinfo {title} {Challenges in
  {{Inferring}} the {{Directionality}} of {{Active Molecular Processes}} from
  {{Single-Molecule Fluorescence Resonance Energy Transfer Trajectories}}},\
  }\href {https://doi.org/10.1021/acs.jpclett.2c03244} {\bibfield  {journal}
  {\bibinfo  {journal} {The Journal of Physical Chemistry Letters}\ }\textbf
  {\bibinfo {volume} {14}},\ \bibinfo {pages} {49} (\bibinfo {year}
  {2023})}\BibitemShut {NoStop}%
\bibitem [{\citenamefont {Igoshin}\ \emph
  {et~al.}(2025{\natexlab{b}})\citenamefont {Igoshin}, \citenamefont
  {Kolomeisky},\ and\ \citenamefont {Makarov}}]{igos25}%
  \BibitemOpen
  \bibfield  {author} {\bibinfo {author} {\bibfnamefont {O.~A.}\ \bibnamefont
  {Igoshin}}, \bibinfo {author} {\bibfnamefont {A.~B.}\ \bibnamefont
  {Kolomeisky}},\ and\ \bibinfo {author} {\bibfnamefont {D.~E.}\ \bibnamefont
  {Makarov}},\ }\bibfield  {title} {\bibinfo {title} {Uncovering dissipation
  from coarse observables: {{A}} case study of a random walk with unobserved
  internal states},\ }\href {https://doi.org/10.1063/5.0247331} {\bibfield
  {journal} {\bibinfo  {journal} {J.\ Chem.\ Phys.}\ }\textbf {\bibinfo
  {volume} {162}},\ \bibinfo {pages} {034111} (\bibinfo {year}
  {2025}{\natexlab{b}})}\BibitemShut {NoStop}%
\bibitem [{\citenamefont {Li}\ \emph {et~al.}(2014)\citenamefont {Li},
  \citenamefont {Kolomeisky},\ and\ \citenamefont {Valleriani}}]{li14}%
  \BibitemOpen
  \bibfield  {author} {\bibinfo {author} {\bibfnamefont {X.}~\bibnamefont
  {Li}}, \bibinfo {author} {\bibfnamefont {A.~B.}\ \bibnamefont {Kolomeisky}},\
  and\ \bibinfo {author} {\bibfnamefont {A.}~\bibnamefont {Valleriani}},\
  }\bibfield  {title} {\bibinfo {title} {{Pathway structure determination in
  complex stochastic networks with non-exponential dwell times}},\ }\href
  {https://doi.org/10.1063/1.4874113} {\bibfield  {journal} {\bibinfo
  {journal} {J.\ Chem.\ Phys.}\ }\textbf {\bibinfo {volume} {140}},\ \bibinfo
  {pages} {184102} (\bibinfo {year} {2014})}\BibitemShut {NoStop}%
\bibitem [{\citenamefont {Sekimoto}(2022)}]{seki22}%
  \BibitemOpen
  \bibfield  {author} {\bibinfo {author} {\bibfnamefont {K.}~\bibnamefont
  {Sekimoto}},\ }\bibfield  {title} {\bibinfo {title} {Derivation of the first
  passage time distribution for markovian process on discrete network},\
  }\href@noop {} {\  (\bibinfo {year} {2022})},\ \Eprint
  {https://arxiv.org/abs/2110.02216} {arXiv:2110.02216 [cond-mat.stat-mech]}
  \BibitemShut {NoStop}%
\end{thebibliography}%
\onecolumngrid

\newpage

\section{Proof of the bounds \eqref{eq:sec3:aff_result}}
\label{sec:app:affinity_proof}

In this section, we prove the bound \eqref{eq:sec3:aff_result} in \Cref{sec:INPpbeEstimation} of the main text. The proof follows a strategy similar to related results in Refs. \cite{vdm22, degu24}. To ensure a self-contained presentation, we summarize the steps in the proof that are identical to the reasoning in Ref. \cite{degu24}, but refer to there for more details.

We assume an underlying Markov network in a stationary state, in which several transitions are observed as described in the main text. Let $\gamma$ be a microscopic trajectory that starts and ends with observed transitions $I, J$, respectively, contains no other visible transitions and has a total duration $t$. The summation over the path weight of all possible trajectories of this form yields the waiting-time distribution \cite{vdm22}
\begin{equation} \label{eq:app1:psidef}
\Psi_{I \to J}(t) = \sum_{\substack{\gamma \text{ from } I \text{ to } J, \\ \text{length } t}} \mathcal{P}[\gamma|I]
\end{equation}
A sum over the corresponding time-reverse paths $\tilde{\gamma}$ yields the according waiting-time distribution $\Psi_{\tilde{J} \to \tilde{I}}(t)$. The crucial idea of the proof is to introduce a different involution $\mathcal{R}: \gamma \mapsto \mathcal{R} \gamma$ which bijectively maps paths from $I$ to $J$ of length $t$ to paths from $\rev{J}$ to $\rev{I}$ with the same duration $t$, so that
\begin{equation}
\label{eq:app1:P1} 
    \Psi_{\tilde{J} \to \tilde{I}}(t) = \sum_{\substack{\gamma \text{ from } \tilde{J} \text{ to } \tilde{I}, \\ \text{length } t}} \mathcal{P}[\gamma|\tilde{J}] = \sum_{\substack{\gamma \text{ from } I \text{ to } J, \\ \text{length } t}} \mathcal{P}[\mathcal{R} \gamma|\tilde{J}]
.\end{equation}
In this case, we specify $\mathcal{R}$ as the ordinary time reversal whenever $\gamma$ is a trimmed path, i.e., whenever $\gamma$ does not contain any cycles. If the unobserved part of $\gamma$ between $I$ and $J$ visits any state more than once, i.e., if $\gamma$ contains cycles, these cycles are \emph{not} reversed, but are traveled in the same direction for both $\gamma$ and $\mathcal{R} \gamma$. For example, the trajectories
\begin{alignat}{10}
\gamma = \, \overset{I}{\to} 4 \to 3 & \to 2 \to 4 \to 3 && \to 8 \to 6 \to 8 && \to 9 \overset{J}{\to} \text{ and } \nonumber \\
\hat{\gamma} = \, \overset{I}{\to} 4 \to 3 & && \to 8 && \to 9 \overset{J}{\to}
\end{alignat}
get mapped to
\begin{alignat}{10}
\mathcal{R} \hat{\gamma} = \, \overset{\tilde{J}}{\to} 9 \to 8 & && \to 3 && \to4 \overset{\tilde{I}}{\to} \text{ and } \nonumber \\
\mathcal{R} \gamma = \, \overset{\tilde{J}}{\to} 9 \to 8 & \to 6 \to 8 && \to 3 \to 2 \to 4 \to 3 && \to 4 \overset{\tilde{I}}{\to}
,\end{alignat}
which clarifies that $\mathcal{R}$ can be understood as ``taking out all cycles, reversing the path, then restoring all cycles in original direction''. The operation $\mathcal{R}$ does not affect residence times in a state, and therefore does not change the duration of a trajectory.

We now introduce the quantity 
\begin{equation} \label{eq:app1:A}
    A[\gamma] \equiv \ln \frac{P(I)}{P(J)} + \ln\frac{\Path{\gamma|I}}{\Path{\R\gamma|\rev{J}}}
.\end{equation}
This quantity bears some resemblance to entropy production, e.g., it satisfies a fluctuation relation $\braket{\exp (-A)} = 1$ and is identical to the microscopic entropy production $\Delta s[\gamma]$ if $\gamma$ does not contain any cycles between $I$ and $J$ because in this case $\mathcal{R}$ coincides with time reversal. We combine \Cref{eq:app1:A} and \Cref{eq:app1:P1} to write
\begin{equation} \label{eq:app1:P2}
        P(I) \Psi_{I \to J}(t)  = P(I) \sum_{\gamma|I \overset{t}{\to} J} \Path{\gamma|I} = P(J) \sum_{\gamma|I \overset{t}{\to} J} \Path{\mathcal{R} \gamma|\widetilde{J}} e^{A[\gamma]}
,\end{equation}
where the notation $I \overset{t}{\to} J$ indicates that summation is over paths $\gamma$ from $I$ to $J$ of length $t$. Combining the definition of $\spath{IJ}{t}$ in \eqref{eq:defaIJ} with Equations \eqref{eq:app1:P1} and \eqref{eq:app1:P2}, we obtain
\begin{equation} \label{eq:app1:P3}
    e^{\spath{IJ}{t}} =\frac{P(I) \Psi_{I \to J}(t)}{P(J) \Psi_{\tilde{J} \to \tilde{I}}(t)} = \frac{\sum_{\gamma|I \overset{t}{\to} J} \Path{\mathcal{R} \gamma|\widetilde{J}} e^{A[\gamma]}}{\sum_{\gamma|I \overset{t}{\to} J} \Path{\mathcal{R} \gamma|\widetilde{J}}} = \mean{e^{A[\gamma]}}_\text{aux},
\end{equation}
and in the last equality identify the expression as a mean with respect to an auxiliary probability measure. Thus, $\inf_{\gamma|I \overset{t}{\to} J} e^{A[\gamma]} \leq \mean{\exp A[\gamma]}_\text{aux} \leq \sup_{\gamma|I \overset{t}{\to} J} e^{A[\gamma]}$, which is equivalent to
\begin{equation} \label{eq:app1:P5}
    \inf_{\gamma|I \overset{t}{\to} J} A[\gamma] \leq \spath{IJ}{t} \leq \sup_{\gamma|I \overset{t}{\to} J} A[\gamma]
.\end{equation}
To recast this relation into the form presented in the main text, we now relate the upper and lower bound on $\spath{IJ}{t}$ to the microscopic entropy production more explicitly. We first note that for the observed transitions $I = k \to l$ and $J = o \to p$ all paths $\gamma$ are of the form $k \to l \to \cdots \to o \to p$. 

In a first step, we assume that $\gamma$ is a trimmed path, which implies $\mathcal{R} \gamma = \tilde{\gamma}$. In this case, the formal calculation
\begin{align}
    A[\gamma] & = \ln \frac{P(I)}{P(J)} + \ln\frac{\Path{\gamma|I}}{\Path{\R\gamma|\rev{J}}} = \ln \frac{p^s_k k_{kl}}{p^s_ok_{op}} + \ln\frac{\Path{\gamma|l}}{\Path{\R\gamma|o}} \\
    & = \ln\frac{\Path{\gamma, \text{ excluding } o \to p}}{\Path{\widetilde{\gamma}, \text{ excluding } p \to o}} = \Delta s[\gamma]
\end{align}
establishes that $A$ coincides with the physical entropy production excluding the final transition. For a general $\gamma$ that may include hidden cycles, we replace the second line in the previous calculation with 
\begin{align}
    A[\gamma] = \ln\frac{\Path{\hat{\gamma}, \text{ excluding } o \to p}}{\Path{\widetilde{\hat{\gamma}}, \text{ excluding } p \to o}} = \Delta s[\hat{\gamma}]
,\end{align}
which makes use of the fact that $\mathcal{R}$ acts like time reversal on the trimmed path, but does not reverse the direction of cycles, so that their contribution to the path weight in the denominator cancels with a corresponding term in the numerator.

As a consequence of the identification $A[\gamma] = \Delta s[\hat{\gamma}]$, we also see that the upper and lower bound in \eqref{eq:app1:P5} are independent of $t$, because residence times within states contribute to the time-symmetric part of a path weight and therefore do not affect entropy production. Therefore, the same bounds hold true for all $t$ in the inequalities \eqref{eq:app1:P5},  allowing the reformulation
\begin{equation}
    \min_{\gamma|I \to J} \Delta s[\hat{\gamma}] \leq \inf_t \spath{IJ}{t} \leq \sup_t \spath{IJ}{t} \leq \max_{\gamma|I \to J} \Delta s[\hat{\gamma}] 
    \label{eq:app1:aff_result}
,\end{equation}
which appears as inequality \eqref{eq:sec3:aff_result} in the main text. We note that ``inf'' and ``sup'' from \eqref{eq:app1:P5} have been replaced by ``min'' and ``max'', respectively, because there are only finitely many distinct values of $\Delta s[\hat{\gamma}]$ corresponding to the different possible trimmed paths from $I$ to $J$.

\newpage
\section{General method to reconstruct realizations of a minimal graph}\label{sec:RGGalgo}

Based on the refined approach to reconstruct a minimal graph illustrated in \Cref{sec:RGGpba}, we present
an algorithm for arbitrary Markov networks, which differs slightly from the simplified flowchart \ref{fig:flowchart} in \Cref{sec:RGGalgoflow}. For a general network with a set of $\NiIJ{1}{\dots}$ and
$\obsNtwo{\dots} + \NiIJ{1}{\dots}$, the following pseudocode describes an algorithm designed to yield the minimal graph if it
is unique or all of its putative realizations otherwise.
\begin{center}
\begin{tabular}{rl}
\hline
\multicolumn{2}{l}{\textbf{Algorithm} \hypertarget{algodMG}{drawMinimalGraph$(\NiIJ{1}{\dots},\obsNtwo{\dots})$}}\\ \hline
{\color{green}A1:} & \begin{minipage}[t]{\linewidth-2.85em} \ttfamily Determine for all pairs in $\left\lbrace I_\pm,J_\pm,K_\pm,\dots\right\rbrace$ whether or not they lie in the same putative cluster (PC) \end{minipage} \\
{\color{green}A2:} & \begin{minipage}[t]{\linewidth-2.85em} $N_\text{Cl}\correspondshat$ \texttt{max. number of clusters;} $\mathcal{C}=\left\lbrace \lbrace I_\pm,L_\pm,\dots\rbrace,\lbrace J_\pm,\dots\rbrace,\dots\right\rbrace\correspondshat$ \texttt{set of sets of visible links lying in the same of the} $N_\text{Cl}$ \ttfamily PCs \end{minipage} \\
{\color{blue}\hypertarget{dMGbi}{B1}:} & $n\coloneqq N_\text{Cl}$, $x\coloneqq 0$, $y\coloneqq 2$ \\
{\color{blue}B2:} & \textbf{Do} \\
{\color{blue}B3:} & \hskip2em\begin{minipage}[t]{\linewidth-4.85em} \ttfamily Select [one of] the pair[s] of $x$ drawn and $y$ not yet drawn links $A_\pm,B_\pm$ with smallest $\NiIJ{1}{A_aB_b}\ (a,b\in\lbrace +,-\rbrace)$ [that lie for $N_\text{Cl} > 1$ in distinct PCs] \end{minipage} \\
{\color{blue}B4:} & \hskip2em\begin{minipage}[t]{\linewidth-4.15em} \ttfamily Draw the undrawn link(s) $(A_\pm,)B_\pm$ and [one of] the shortest path[s] connecting $A_\pm$ and $B_\pm$. If there are several distinct realizations thereof, generate a separate graph for each of them. \end{minipage} \\
{\color{blue}B5:} &\hskip2em $n\coloneqq n-1$, $x\coloneqq 1$, $y\coloneqq 1$ \\
{\color{blue}B6:} & \textbf{while} $n>1$ \\
{C1:} & \textbf{Do} \\
{C2:} &\ttfamily \hskip2em\begin{minipage}[t]{\linewidth-4.85em} Select an unlabeled set from $\mathcal{C}$, the PC of which is connected to the largest number of other PCs, and label it $c_n$ \end{minipage} \\
{C3:} & \hskip2em\textcolor{blue}{$f_1\left(c_n, \text{\emph{"shortest"}}, \text{\emph{"the same"}} \right)$} \\
{C4:} & \hskip2em$n\coloneqq n+1$ \\
{C5:} & \textbf{while} $n\leq N_\text{Cl}$ \\
{C6:} &\ttfamily Delete labels $c_n$ \\
{C7:} & \textcolor{blue}{$f_2\left(N_\text{Cl}, \mathcal{C}, \text{\emph{"shortest"}}\right)$} \\
{C8:} & \begin{minipage}[t]{\linewidth-2.85em} \textbf{If} C7 \texttt{returns realizations with updated} $N_\text{Cl}$ \texttt{and} $\mathcal{C}$:\end{minipage} \\
{C9:} & \hskip2em\begin{minipage}[t]{\linewidth-4.85em} \ttfamily Discard all realizations with updated {\normalfont$N_\text{Cl}$} and $\mathcal{C}$, and restart from line {\normalfont B1} for each of the updated combinations {\normalfont$N_\text{Cl},\mathcal{C}$} \end{minipage} \\
{C10:} & \textbf{end if} \\
{\color{blue}D1:} & $n\coloneqq 1$ \\
{\color{blue}D2:} & \textbf{Do} \\
{\color{blue}D3:} & \hskip2em\begin{minipage}[t]{\linewidth-4.84em} \ttfamily From $\mathcal{C}$, select an unlabeled set, the PC of which is connected to the largest number of other PCs, and label it $c_n$ \end{minipage} \\
{\color{blue}D4:} & \hskip2em\textcolor{blue}{$f_1\left(c_n, \text{\emph{"second-shortest"}}, \text{\emph{"the same"}} \right)$} \\
{\color{blue}D5:} & \hskip2em$n\coloneqq n+1$ \\
{\color{blue}D6:} & \textbf{while} $n\leq N_\text{Cl}$ \\
{\color{blue}D7:} &\ttfamily Delete labels $c_n$ \\
{\color{blue}D8:} & \textcolor{blue}{$f_2\left(N_\text{Cl}, \mathcal{C}, \text{\emph{"second-shortest"}}\right)$} \\
{\color{blue}D9:} & \begin{minipage}[t]{\linewidth-2.85em} \textbf{If} D8 \texttt{returns realizations with updated} $N_\text{Cl}$ \texttt{and} $\mathcal{C}$: \end{minipage} \\
{\color{blue}D10:} & \hskip2em\begin{minipage}[t]{\linewidth-4.85em} \ttfamily Discard all realizations with updated {\normalfont$N_\text{Cl}$} and $\mathcal{C}$, and restart from line {\normalfont B1} for each of the updated combinations {\normalfont$N_\text{Cl},\mathcal{C}$} \end{minipage} \\
{\color{blue}D11:} & \textbf{end if}
\end{tabular}
\end{center}
The algorithm \hyperlink{algodMG}{drawMinimalGraph$(\NiIJ{1}{\dots},\obsNtwo{\dots})$} consists of four logical blocks A through D, in which
lines (instructions) are numbered and indentations highlight instructions within loops and conditionals.
Additionally, we highlight that the algorithm may generate multiple putative realizations in separate graphs. The
remaining steps of the algorithm then need to be applied to these graphs separately. \par
The first block (A) makes use of the pair rule, which we have discussed in \Cref{sec:INPwsfclusters}, to identify the
maximum number $N_\text{Cl}$ of putative clusters (PCs) and the set $\mathcal{C}$ containing sets of visible links lying in
the same of the $N_\text{Cl}$ PCs.
Block B uses these intermediate results to connect one visible link from each PC to one of each neighboring PC.
When selecting links to draw, there might be sequences of visible links with equal $\NiIJ{1}{\dots}$. In these cases
and all similar cases in later blocks, we arbitrarily choose one of the corresponding paths. For block B, this leads to
a tree-like graph, which consists of one visible link of each PC with a
shortest path to a visible link of each neighboring PC.\par
The loop in block C generates realizations with shortest paths between all visible links by consecutively invoking
functions $f_1$ and $f_2$. These functions both depend on $\NiIJ{1}{\dots}$ and $\obsNtwo{\dots}$, which we omit in our notation for the sake of brevity.
Function \hyperlink{algofi}{$f_1\left(c_n, \textit{"shortest"}, \textit{"the same"}\right)$} generates, with its default arguments
\textit{"shortest"} and \textit{"the same"}, in all putative realizations all shortest paths within the PC that
corresponds to the set $c_n\in\mathcal{C}$ as follows.
\begin{center}
\begin{tabular}{rl}
\hline
\multicolumn{2}{l}{\textbf{Algorithm} \hypertarget{algofi}{$f_1\left(c_n, \textit{"shortest"}, \textit{"the same"}\right)$}}\\ \hline
A1: & $m\coloneqq 1$ \\
A2: & \textbf{Do} \\
A3: & \hskip2em\begin{minipage}[t]{\linewidth-5.3em} \ttfamily Select (a yet unmarked) $L_\pm\in c_n$ with [one of] the {\normalfont\textit{shortest}} path[s], from $L_l$ to $I_i$ with $l,i\in\lbrace+,-\rbrace$, for an already drawn $I_\pm$ in {\normalfont\textit{the same}} PC (where $L_\pm=I_\pm$ due to $\NiIJ{1}{I_+I_-}=0$ is allowed in the case of {\normalfont\textit{"the same"}}). Mark $L_\pm$ \end{minipage} \\
\hypertarget{fiaiv}{A4}: & \hskip2em\begin{minipage}[t]{\linewidth-5.3em} \ttfamily Draw $L_\pm$ if it is not yet drawn and the {\normalfont\textit{shortest}} hidden paths between $L_\pm$ and any already drawn visible link in {\normalfont\textit{the same}} PC. If there are several distinct realizations thereof, generate a separate graph for each of them. [For the case of {\normalfont\textit{"second-shortest"}}, mind the initial ambiguity of $\obsNtwo{L_lI_i}=1$ discussed in \Cref{sec:shortestpathsL}] \end{minipage} \\
A5: & \hskip2em\begin{minipage}[t]{\linewidth-5.3em} \ttfamily Check each realization for respecting all $\NiIJ{1}{\dots}$ and $\obsNtwo{\dots}$ of drawn links and discard those that do not. \end{minipage} \\
A6: & \hskip2em$m\coloneqq m+1$ \\
A7: & \textbf{while} $m\leq\abs{c_n}$ \\
A8: & \ttfamily Unmark visible links \\
E1: & \begin{minipage}[t]{\linewidth-3.3em} \textbf{Result:} \texttt{Putative realizations with} \textit{shortest} \ttfamily paths between all links in $c_n$ \end{minipage}
\end{tabular}
\end{center}
For the default arguments, line \hyperlink{fiaiv}{A4} needs to be executed in a way that may change which links and states form a bridge but keeps
neighboring PCs connected by a bridge. Moreover, we stress that putative realizations resulting from this line must
respect the relevant $\MiIJK{1}{L_lI_i}{K}$ introduced in \Cref{sec:shortestpathsL} for the smallest set $K$ of visible
transitions that are considered to be hidden.

Function \hyperlink{algofii}{$f_2\left(N_\text{Cl}, \mathcal{C}, \textit{"shortest"}\right)$} generates, for its default argument
\textit{"shortest"}, shortest paths between PCs beginning in the tree of PCs with the outer ones as follows.
\begin{center}
\begin{tabular}{rl}
\hline
\multicolumn{2}{l}{\textbf{Algorithm} \hypertarget{algofii}{$f_2\left(N_\text{Cl}, \mathcal{C}, \textit{"shortest"}\right)$}} \\ \hline
A1: & $m\coloneqq 1$ \\
A2: & \textbf{Do} \\
A3: & \hskip2em\begin{minipage}[t]{\linewidth-4.37em} \ttfamily Select one of the unlabeled PCs with the smallest amount of neighboring PCs in the tree of PCs. Label it $m$ and its set in $\mathcal{C}$ as $c_m$. \end{minipage} \\
\hypertarget{fiiaiv}{A4}: & \hskip2em\begin{minipage}[t]{\linewidth-5.3em} \textcolor{blue}{$f_1\left(c_m, \textit{"shortest"}, \textit{"a distinct"}\right)$} \end{minipage} \\
E1: & \hskip2em\begin{minipage}[t]{\linewidth-5.3em} \textbf{If} $N_\text{Cl}$ \ttfamily decreases for a putative realization:\end{minipage} \\
E2: & \hskip3em\begin{minipage}[t]{\linewidth-5.37em} \ttfamily Update {\normalfont$N_\text{Cl}$} and $\mathcal{C}$ only for all of these putative realizations separately \end{minipage} \\
\hypertarget{fiieiii}{E3}: & \hskip3em\begin{minipage}[t]{\linewidth-5.37em} \ttfamily Exit this function for these putative realizations \end{minipage} \\
E4: & \hskip2em \textbf{end if} \\
A6: & \hskip2em$m\coloneqq m+1$ \\
A7: & \textbf{while} $m\leq N_\text{Cl}$ \\
E5: & \begin{minipage}[t]{\linewidth-3.3em} \textbf{Result:} \texttt{Putative realizations with} \textit{shortest} \texttt{paths} \end{minipage}
\end{tabular}
\end{center}
For each iteration, line \hyperlink{fiiaiv}{A4} may lead to modified bridges or merged PCs. Thereby, either
$N_\text{Cl}$ and $\mathcal{C}$ remain unchanged for a realization or they change due to merging two or more PCs by
drawing a path between two PCs that contains not all of the bridges formerly connecting them. In
these latter cases, the function $f_2$ exits via line \hyperlink{fiieiii}{E3}, whereupon the algorithm
\hyperlink{algodMG}{drawMinimalGraph$(\NiIJ{1}{\dots},\obsNtwo{\dots})$} proceeds with deleting the corresponding graph
and restarting with the updated quantities in line \hyperlink{dMGbi}{B1} because finding all putative new realizations
is crucial.

Block D is the analogous procedure focusing on second-shortest paths. Thereafter, additional information or methods
might allow discarding realizations and extending others with indications of yet undrawn subgraphs.
Such methods may be based on specific characteristics of the graph between
visible transitions, examples of which we present in \Cref{sec:RGGextFgraphs}.
In the end, the remaining realizations contain shortest and second-shortest self-avoiding
paths. In particular, they are all realizations of the minimal graph that are compatible with the information used about
the given nonequilibrium process. If only one realization results, it is the minimal graph.

Note that this algorithm is designed to provide a supporting structure to find all putative realizations and is not verified to be optimal.
As an example, deleting existing realizations for which $N_\text{Cl}$
and $\mathcal{C}$ have changed is not strictly necessary in lines C9 and D10. Without this, we would then jump to line C1 setting $n=1$. In this block (C),
we would however need to generate putative alternatives for existing paths between visible links, which is less straightforward
to describe unambiguously but may speed up the algorithm. Similarly, another route could be adding and completing cycles before
drawing paths between distinct visible links. Yet, selecting an approach is like choosing between high road and low road---only
at the end we know which route is faster.

\newpage

\section{Additional methods for graph reconstructions}\label{sec:RGGextFgraphs}
\subsection{Cyclic rate product and steady state condition}
Rates of visible transitions $I$ as well as of hidden transitions along a path $i\overset{I}{\to}j\to k
\overset{J}{\to} l$ with $\NiIJ{1}{IJ} = 1$ follow from the short-time limit $t\to 0$ as
\begin{align}
k_{ij} &= \lim_{t\to 0}\wtdPsi{\tilde{I}}{I}{t} \label{eq:defkijVIS} \qq{and}\\
k_{jk} &= \lim_{t\to 0}\wtdPsi{I}{J}{t}/\left(tk_{kl}\right), \label{eq:defkjk}
\end{align}
respectively, since the product of transition rates of the shortest path between $I$ and $J$ dominates
$\wtdPsi{I}{J}{t}$ in this limit \cite{maie24}. As shorthand, we define the set $\mathcal{V}_i$
to contain all transitions for which the rates are inferable. These rates allow us to infer additional
information about hidden edges near visible ones in specific situations we will discuss now.

First, consider inferred rates $k_{jj+1}$ of a cycle that corresponds to $\NiIJ{1}{II} = n$. An inequality in
\begin{align}
\ln(\prod_{j=1}^n k_{jj+1}) \overset{?}{=} \lim_{t\to 0} \ln\left(\wtdPsi{I}{I}{t} \frac{n!}{t^n} \right)
\label{eq:conditionCycle}
\end{align}
reveals the existence of at least one more cycle with $\NiIJ{1}{II} = n$. 

Hence, this test may provide insight when $\obsNtwo{IJ}=1$ so that, as discussed in \Cref{sec:shortestpathsL}, a degenerate shortest path is possible. If we are able to determine the transition rates via Eqs. \eqref{eq:defkijVIS} and \eqref{eq:defkjk}, both sides of Eq.\eqref{eq:conditionCycle} can be determined and compared. If the results do not match, we know for certain that the shortest path is indeed degenerate. We illustrate this point with a network from the main text, namely the one shown in \Cref{fig:graphExample}. We assume that
only the number of hidden transitions along shortest paths have been inferred. For the cycle $1\to 2\to 3\to 1$,
\Cref{eq:defkijVIS,eq:defkjk} yield all rates $k_{jj+1}$. Furthermore, comparison \eqref{eq:conditionCycle} becomes an
inequality due to the cycle $1\to 2\to 6\to 1$ containing $R_\pm$ and $\NiIJ{1}{R_+R_+} = 2$ hidden transitions, and is therefore able to detect this additional cycle.

Second, for a transition $I=(ij)$, its mean observed rate $\meanNness{I}$ and the rate $k_{ij}$ determine the steady-state
probability $\psteady{i}$ by $\psteady{i} = \meanNness{I}/k_{ij}$ \cite{maie24}. An inequality in
\begin{align}
\partial_t\psteady{i} = 0 \overset{?}{=} \sum_{ij\in\mathcal{V}_i}(\psteady{i}k_{ij}-\psteady{j}k_{ji}) = \sum_{ij\in\mathcal{V}_i}\jsteady{ij}, \label{eq:conditionState}
\end{align}
where we sum over the set $\mathcal{V}_i$ of known, inferable net currents involving state $i$ on the right-hand side, suggests the existence of additional
hidden edges that begin/end at state $i$.

An alternative uses the escape rate
\begin{align}
\Gamma_i = -\lim_{t\to 0}\dv{t}\ln\frac{\wtdPsi{\tilde{I}}{I}{t}}{k_{ij}} \label{eq:escapeRate}
\end{align}
of state $i$, which dominates the exponential decay of $\wtdPsi{\tilde{I}}{I}{t}$ in the short-time limit.
Since this rate equals the sum over all rates of transitions leaving state $i$, an inequality in
\begin{align}
\Gamma_i \overset{?}{=} \sum_{ij\in\mathcal{V}_i} k_{ij} \label{eq:conditionStateERate}
\end{align}
again implies at least one additional existing hidden edge at state $i$.
In the example discussed in \Cref{sec:RGG_cycle_based_app}, where we reconstruct the graph shown in \Cref{fig:graphExample}, finding an inequality in
comparison \eqref{eq:conditionState} or \eqref{eq:conditionStateERate} for state 3 would mean that the true graph includes
at least one more edge starting at state 3.
Such a finding indicates that the minimal graph is not the full one and provides a starting point fur further inference, which can be marked in the minimal graph(s).

\subsection{Entropy production along paths and affinities} \label{sec:RMGFTepra}
For a Markov network in a NESS with only one self-avoiding path between two visible links $I_\pm,J_\pm$, as is the case in, e.g.,
\Cref{fig:rggPBAexample}\,a), entropy production along each observed trajectory is constant.
Consequently and in analogy to the cycle affinity discussed in Ref. \cite{vdm22}, the quantity
\begin{align}
\Psiaff{I_+J_+}{t} = \PsiaffLn{I_+}{J_+}{J_-}{I_-}
\end{align}
is constant
and reduces to the logarithmic ratio of rates along the path of $I_+\to J_+$ and the path of $J_-\to I_-$, which, e.g., yields
$\Psiaff{L_+R_+}{t} = a_{L_+R_+} = \ln(k_{45}k_{56}k_{67}/k_{65}k_{54}k_{41})$ for the graph shown in \Cref{fig:rggPBAexample}\,a).
Moreover, we have
\begin{align}
\Psiaff{L_+R_+}{t} + \Psiaff{R_+L_-}{t} &= a_{L_+R_+} + a_{R_+L_-} = \ln\frac{k_{45}k_{56}k_{67}}{k_{65}k_{54}k_{41}}
+ \ln\frac{k_{75}k_{54}k_{41}}{k_{45}k_{57}k_{76}} \notag \\
&= \ln\frac{k_{56}k_{67}k_{75}}{k_{65}k_{57}k_{76}} = a_{R_+R_+}= \Psiaff{R_+R_+}{t} \label{eq:epralongpathBSP}
\end{align}
in this example. We note that the quantity \eqref{eq:epralongpathBSP} differs from $\spath{\dots}{t}$ as introduced in the main text in definition \eqref{eq:defaIJ} only by an additional term $\ln(P(R_-)/P(R_+))$. 

Given a candidate for a minimal graph like the one in \Cref{fig:rggPBAexample}\,c), such consistency checks provide additional evidence supporting that the minimal graph is correct. In the general situation, explicitly verifying equalities like \Cref{eq:epralongpathBSP} provide a necessary criterion only, because in topologies with degenerate shortest paths relations like \eqref{eq:epralongpathBSP} and \eqref{eq:conditionCycle} are not valid. Under the additional conditions that there are no degenerate shortest paths and no additional self-avoiding paths with the same entropy production, however, consistency checks like \eqref{eq:epralongpathBSP} provide another necessary and sufficient criterion to conclude uniqueness of a hidden path between two transitions, in addition to constancy of the quantities $a_{\dots}(t) = \text{const.}$ themselves.

For the graph in \Cref{fig:graphExample} for example, the constant $\spath{L_+R_-}{t}$ leads to $\obsNtwo{L_+R_-}=0$,
which means that we do not find an additional path between $L_+$ and $R_-$ with entropy production differing from the one along the shortest path
(cf. \Cref{sec:shortestpathsL}). In this example the minimal graph is consistent with both constancy of the $\spath{\dots}{t}$ in time and the consistency check \eqref{eq:epralongpathBSP}. 
As a consequence, the minimal graph has no additional self-avoiding paths between states 2 and 4, which is in agreement with the true graph. In this sense, this consistency check provides a restriction on both the minimal graph and possible extensions.

\newpage
\section{Model parameters and simulation}
\subsection{Multicyclic 10-state network for the illustration of topological inference in \Cref{fig:setupExamples}}\label{app:MPSeggtopo}
The multicyclic 10-state network in \Cref{fig:setupExamples} contains 10 hidden states, 12 hidden links and 1 visible pair of
transitions. The transition rates given in \Cref{tab:paramsSetup} lead to the waiting-time
distributions that are shown in \Cref{fig:setupExamples}\,b) and c) and used for the logarithmic ratio
$\spath{++}{t}$ as defined in \Cref{eq:defaIJ} and illustrated in \Cref{fig:setupExamples}\,d).
The waiting-time distributions can be derived from the solution of initial value problems based on the absorbing
network, in which observed transitions are redirected into auxiliary states \cite{vdm22,seki22}.
\begin{table}[H]
\centering
\caption{Transition rates of the multicyclic 10-state network in \Cref{fig:setupExamples}.}
\label{tab:paramsSetup}
\begin{tabular}{|c|c|c|c|c|c|c|c|c|c|}
\hline
State 1 & State 2 & State 3 & State 4 & State 5 & State 6 & State 7 & State 8 & State 9 & State 10 \\ \hline
$k_{12} = 1.0$  & $k_{21} = 1.0$ & $k_{32} = 1.0$ & $k_{42} = 2.5$ & $k_{54} = 3.0$ & $k_{65} = 1.0$ & $k_{74} = 1.0$ & $k_{83} = 1.0$ & $k_{98} = 1.0$  & $k_{101} = 2.0$ \\
$k_{110} = 2.0$ & $k_{23} = 1.0$ & $k_{34} = 1.0$ & $k_{43} = 3.0$ & $k_{56} = 1.0$ & $k_{67} = 4.0$ & $k_{76} = 1.0$ & $k_{86} = 8.0$ & $k_{910} = 4.0$ & $k_{109} = 1.0$ \\
&                 $k_{24} = 1.0$ & $k_{38} = 0.2$ & $k_{45} = 3.0$ &                & $k_{68} = 1.0$ &                & $k_{89} = 3.0$ &                 & \\
&                                &                & $k_{47} = 1.0$ &                &                &                &                &                 & \\ \hline
\end{tabular}
\end{table}
\subsection{Multicyclic 10-state network for the illustration of entropy production along paths in \Cref{fig:PBEEexample}}\label{app:MPSeggpathepr}
The graph of the multicyclic 10-state network in \Cref{fig:PBEEexample} differs from the graph of the network in
\Cref{fig:setupExamples} by its visible links $I_\pm$ and $J_\pm$. The transition rates given in
\Cref{tab:paramsPBEE} determine the NESS of the network whose entropy production along trajectories from $I_+$ to
$J_+$ excluding the last transition is shown in \Cref{fig:PBEEexample}\,b). The microscopic entropy production
along the three paths from $I_+$ to $J_+$ directly results from the transition rates of their forward and backward direction.
The coarse-grained entropy production can be derived from the rates, the stationary state and the waiting-time
distributions, which are determined as described in the previous \Cref{app:MPSeggtopo}.

For the $1 731 699$ networks for which the scatter plot in \Cref{fig:PBEEexample}\,c) displays the ratio of differences between microscopic and
between coarse-grained entropy productions along trajectories from $I_+$ to $J_+$, we have randomly drawn all transition rates listed in
\Cref{tab:paramsPBEE} from the uniform distribution $\Theta(0.5,10)$ that is defined between $0.5$ and $10$.
The binned mean of the shown data has been determined in bins of width $0.1$ for $\Delta s_{\to 4\to 3\to}\in[-4,4)$,
in bins of width $1.0$ for $-6 \leq \Delta s_{\to 4\to 3\to} < 4$ and $4 \leq \Delta s_{\to 4\to 3\to} < 6$, and
in one bin for other values of $\Delta s_{\to 4\to 3\to}$ on either side of zero.
\begin{table}[H]
\centering
\caption{Transition rates of the multicyclic 10-state network used to illustrate coarse-grained and microscopic entropy
production along paths in \Cref{fig:PBEEexample}\,b).}
\label{tab:paramsPBEE}
\begin{tabular}{|c|c|c|c|c|c|c|c|c|c|}
\hline
State 1 & State 2 & State 3 & State 4 & State 5 & State 6 & State 7 & State 8 & State 9 & State 10 \\ \hline 
 $k_{12}=0.45$  & $k_{21}=0.45$ & $k_{32}=16.59$ & $k_{42}=1.41$ & $k_{54}=1.69$ & $k_{65}=47.53$ & $k_{74}=0.45$ & $k_{83}=0.56$ & $k_{98}=0.45$  & $k_{101}=2.25$ \\
 $k_{110}=2.25$ & $k_{23}=16.59$ & $k_{34}=0.84$ & $k_{43}=1.27$ & $k_{56}=237.66$ & $k_{67}=1.80$ & $k_{76}=0.45$ & $k_{86}=4.50$ & $k_{910}=1.80$ & $k_{109}=0.45$ \\
                & $k_{24}=0.56$ & $k_{38}=0.11$ & $k_{45}=1.69$ &               & $k_{68}=0.56$ &               & $k_{89}=1.35$ &  & \\
                &               &               & $k_{47}=0.45$ & & & & & & \\ \hline
\end{tabular}
\end{table}

\end{document}